\documentclass[10pt,letterpaper,reqno]{amsart}
\usepackage{mathrsfs}
\usepackage{amsmath,amsfonts,amssymb,amsthm, color}
\usepackage[T1]{fontenc}
\usepackage{fouridx}
\usepackage[normalem]{ulem}

\usepackage{mathrsfs}
\usepackage{amsmath,amsfonts,amssymb,amsthm, color}
\usepackage[T1]{fontenc}
\usepackage{fouridx}
\usepackage[normalem]{ulem}

\usepackage{graphicx}
\usepackage{amsmath,amssymb}
\usepackage{txfonts}
\usepackage{rotating}
\usepackage[margin=1in,innermargin=1.5in]{geometry}
\usepackage[pdftex]{hyperref}
\usepackage[all]{hypcap}
\usepackage{array}
\usepackage{booktabs}
\usepackage{multirow}
\usepackage[utf8]{inputenc}
\usepackage{fancyhdr}
\usepackage{color}

\theoremstyle{plain}
\newtheorem{theorem}{Theorem}[section]
\theoremstyle{remark}
\newtheorem{remark}[theorem]{Remark}

\theoremstyle{plain}
\newtheorem{corollary}[theorem]{Corollary}
\newtheorem{lemma}[theorem]{Lemma}
\newtheorem{proposition}[theorem]{Proposition}
\newtheorem{definition}[theorem]{Definition}

\numberwithin{equation}{section}

\newcommand\dela[1]{}

\begin{document}

\title[HJMM Equation in the weighted $L^p$ spaces]{Stochastic Evolution Equations in Banach Spaces and Applications to Heath-Jarrow-Morton-Musiela Equation}


\author[Z. Brze\'{z}niak  \and T. Kok]{Zdzis{\l}aw Brze\'{z}niak  \and Tayfun Kok}


\address{
Department of Mathematics, University of York, Heslington, York, YO105DD, UK}
\email{zdzislaw.brzezniak@york.ac.uk, ttk505@york.ac.uk}

\begin{abstract}
In this paper we study the stochastic evolution equation  (\ref{3equa1})  in martingale-type 2 Banach spaces (with the linear part of the drift being only a generator of a $C_0$-semigroup). We prove the existence and the uniqueness of solutions to this equation. We apply the abstract results to the Heath-Jarrow-Morton-Musiela (HJMM) equation (\ref{equationnnnnnn5}). In particular,  we prove the existence and the uniqueness of solutions to the latter equation  in the weighted  Lebesgue  and Sobolev spaces $L^{p}_{\nu}$ and $W^{1,p}_{\nu}$ respectively. We also find a sufficient condition for the existence and the uniqueness of an invariant measure for the Markov semigroup associated to   equation (\ref{equationnnnnnn5}) in the weighted spaces $L^{p}_{\nu}$.
\end{abstract}

\keywords{(SEE)s \and (HJMM) Equation \and mild and strong solutions \and the existence and the uniqueness of solutions \and Markov semigroup \and the existence and the uniqueness of an invariant measure}

\maketitle

\section{Introduction}
\label{s:1}

Let $(\Omega,\mathcal{F},\mathbb{F},\mathbb{P})$, where $\mathbb{F}=\{\mathcal{F}_{t}\}_{t\geq 0}$, be a filtered probability space, and $(H,\langle \cdot,\cdot\rangle_{H})$ be a separable Hilbert space. Assume that $W$ is an $H$-valued, $\mathbb{F}$-cylindrical canonical Wiener process, i.e. $W$ is a family  $\{W(t)\}_{t\geq0}$ of bounded linear operators from $H$ into $L^{2}(\Omega,\mathcal{F},\mathbb{P})$ such that

\noindent (i) for all $t\geq0$ and $h_{1},h_{2}\in H$, $\mathbb{E}W(t)h_{1}W(t)h_{2}=t\langle h_{1},h_{2}\rangle_{H}$,

\noindent (ii) for any $h\in H$, $\{W(t)h\}_{t\geq0}$ is real-valued, $\mathbb{F}$-adapted Wiener process.

The aim of this paper is twofold. The first one is to prove, under the local Lipschitz conditions on the coefficients, the existence and the uniqueness of solutions to the following stochastic evolution equation in a Banach space $X$ satisfying the $H_{p}$ condition, which is stated  below.
\begin{equation}\label{3equa1}
\begin{cases}
du(t)=(Au(t)+F(t,u(t)))dt+G(t,u(t))dW(t),\\
u(0)=u_{0},
\end{cases}
\end{equation}
where $A$ is the generator of  a contraction type, defined later, $C_{0}$ semigroup $\{S(t)\}_{t\geq0}$ on $X$, $u_{0} : \Omega\rightarrow X$ is an $\mathcal{F}_{0}$-measurable, square integrable random variable, and for each $t\geq0$,~~$F(t,\cdot):X\rightarrow X$ and $G(t,\cdot):X\rightarrow\gamma(H,X)$. Here $\gamma(H,X)$ denotes the Banach space of $\gamma$-radonifying operators from $H$ into $X$, see \cite{[Neidhardt]} for details. The norm in the space $\gamma(H,X)$ will be denoted by $\|\cdot\|_{\gamma(H,X)}$. We say that a  Banach space $X$ with the norm $\|\cdot\|_{X}$ satisfies $H_{p}$ condition, see \cite{[Zhu]} for details, if for some $p\geq2$,~~the function $\psi: x\ni X\mapsto \psi(x)=\|x\|_{X}^{p}\in\mathbb{R}$ is of $C^{2}$ class on $X$ (in the $Fr\acute{e}chet$ derivative sense) and there exist constants $K_{1}(p),K_{2}(p)>0$ depending on $p$ such that for every $x\in X$,~~$\left|\psi^{'}(x)\right|\leq K_{1}(p)\|x\|_{X}^{p-1}$ and $\left|\psi^{''}(x)\right|\leq K_{2}(p)\|x\|_{X}^{p-2}$.
\\

The second one is to apply the previous abstract results to the Heath-Jarrow-Morton-Musiela (HJMM) equation (\ref{equationnnnnnn5}) introduced in section \ref{s:3}. The existence and the uniqueness of solutions to equation (\ref{equationnnnnnn5}) in Hilbert spaces, and ergodic properties of the solutions have been studied, in particular, in \cite{[Bjork]}, \cite{[Filipovic4]}, \cite{[Filipovic3]}, \cite{[Filipovic5]}, \cite{[Filipovic1]}, \cite{[Goldys1]}, \cite{[Musiela]}, \cite{[Peszat]}, \cite{[Rusinek2]}, \cite{[Rusinek1]}, \cite{[Vargiolu]}, \cite{[Tehranchi]}. In this paper we prove the existence and the uniqueness of solutions to  equation (\ref{equationnnnnnn5}) in the weighted Banach spaces $L^{p}_{\nu}$ and $W^{1,p}_{\nu}$ defined in section \ref{s:3}. We also find a sufficient condition from Theorem 3.7 in \cite{[brzezniak8]} for the existence and the uniqueness of invariant measures for the solution to  equation (\ref{equationnnnnnn5}), when the coefficients are time independent, in the spaces $L^{p}_{\nu}$. An important feature of our results is that we are able to consider HJMM Equations driven by a cylindrical Wiener process on a (possibly infinite dimensional) Hilbert spaces. For this purpose we use a characterization of $\gamma$-radonifying operators from a Hilbert space to an $L^p$ space found recently by the first named author and Peszat in \cite{[brzezniak3]}.

\section{Mathematical preliminaries}
\label{s:2}

We begin with introducing some notations.

\begin{definition}
A Banach space $X$ with the norm $\|\cdot\|_{X}$ is called a martingale-type 2 Banach space if there exists a constant $C>0$ depending only on $X$ such that for any $X$-valued martingale $\{M_{n}\}_{n\in\mathbb{N}}$, the following inequality holds
\begin{align*}
\sup_{n\in\mathbb{N}}\mathbb{E}\|M_{n}\|_{X}^{2}\leq C\sum_{n}\mathbb{E}\|M_{n}-M_{n-1}\|_{X}^{2}.
\end{align*}
\end{definition}
The Lebesgue function spaces $L^{p}$,~~$p\geq2$,~~are examples of martingale-type 2 Banach spaces.

\begin{proposition}\cite{[Zhu]}
If $X$ is a Banach space satisfying the $H_{p}$ condition, then $X$ is an martingale-type 2 Banach space.
\end{proposition}

\begin{definition}
Let $X$ be a Banach space and $\mathcal{L}(X)$ be the space of all bounded linear operators from $X$ to $X$. A $C_{0}$-semigroup $S=\{S(t)\}_{t\geq 0}$ on $X$ is called contraction type iff there exists a constant $\beta\in\mathbb{R}$ such that
\begin{equation}\label{semigroupp}
\|S(t)\|_{\mathcal{L}(X)}\leq e^{\beta t},\quad t\geq0.
\end{equation}
\end{definition}

\begin{definition}
Let $X$ be a martingale-type 2 Banach space and $S$ be a contraction type $C_{0}$-semigroup on $X$ with the infinitesimal generator $A$. A process $u$ is called an $X$-valued mild solution to SEE (\ref{3equa1}) if for each $t\geq0$,
\begin{equation}\label{3equa2}
u(t)=S(t)u_{0}+\int_{0}^{t}S(t-r)F(r,u(r))dr+\int_{0}^{t}S(t-r)G(r,u(r))dW(r),\quad \mathbb{P}-a.s.
\end{equation}
where $u$ is such that each term on the right hand side is well-defined.
\end{definition}
The integrals in equation (\ref{3equa2}) are not always well-defined for any process $u$, and functions $F$ and $G$. Therefore, we also introduce some notations which make the integrals in equation (\ref{3equa2}) well-defined in martingale type 2 Banach spaces.  Let $X$ be a martingale type 2 Banach space endowed with the norm $\|\cdot\|_{X}$ and $S$ be a contraction type $C_{0}$-semigroup on $X$. It follows from \cite{[brzezniak11]} that if $\xi$ is a  $\gamma(H,X)$-valued, $\mathbb{F}$-progressively measurable process such that for all $t\geq0$,
\begin{align*}
\mathbb{E}\int_{0}^{t}\|\xi(s)\|_{\gamma(H,X)}^{2}ds<\infty,
\end{align*}
then for each $t\geq0$, the stochastic integral $\int_{0}^{t}\xi(s)dW(s)$ is well-defined. Moreover, the stochastic integral process $\int_{0}^{t}\xi(s)dW(s)$, $t\geq0$, is an $X$-valued, $\mathbb{F}$-progressively measurable process, and there exists a constant $C>0$ (independent of $\xi$) such that
\begin{equation}\label{equuu1}
\mathbb{E}\left\|\int_{0}^{t}\xi(s)dW(s)\right\|^{2}_{X}\leq C\int_{0}^{t}\mathbb{E}\|\xi(s)\|^{2}_{\gamma(H,X)}ds,\quad t\geq 0.
\end{equation}
 In particular, for each $t\geq0$, the stochastic convolution integral $\int_{0}^{t}S(t-r)\xi(r)dW(r)$ is well-defined provided that $\xi$ is a $\gamma(H,X)$-valued, $\mathbb{F}$-progressively measurable process such that for each $t\geq0$,
\begin{align*}
\mathbb{E}\int_{0}^{t}\|\xi(r)\|_{\gamma(H,X)}^{2}dr<\infty.
\end{align*}
Moreover, the stochastic convolution process $\int_{0}^{t}S(t-r)\xi(r)dW(r)$,~~$t\geq0$, is an $X$-valued, $\mathbb{F}$ progressively measurable process. If for each $t\geq0$, the function $G(t,\cdot):X\rightarrow \gamma(H,X)$ is bounded, and for each $T>0$ and $x\in X$, the function $G(\cdot,x):[0,T]\rightarrow\gamma(H,X)$ is Borel measurable, then the process $G(t,u(t))$,~~$t\geq0$,~~is $\gamma(H,X)$-valued, $\mathbb{F}$-progressively measurable such that for each $t\geq0$,
\begin{align*}
\mathbb{E}\int_{0}^{t}\|G(r,u(r))\|_{\gamma(H,X)}^{2}dr<\infty
\end{align*}
 provided that $u$ is an $X$-valued, $\mathbb{F}$-progressively measurable process such that for each $t\geq0$,
\begin{align*}
\mathbb{E}\int_{0}^{t}\|u(r)\|_{X}^{2}dr<\infty.
\end{align*}
Hence, for each $t\geq0$, the stochastic convolution integral $\int_{0}^{t}S(t-r)G(r,u(r))dW(r)$ is well-defined, and the stochastic convolution process $\int_{0}^{t}S(t-r)G(r,u(r))dW(r)$,~$t\geq0$, is an $X$-valued, $\mathbb{F}$ progressively measurable process.\\

It follows from \cite{[ball]} that if $u$ is an $X$-valued, $\mathbb{F}$-progressively measurable process such that for each $t\geq0$,
\begin{align*}
\int_{0}^{t}\left\|u(r)\right\|_{X}dr<\infty,\quad \mathbb{P}-a.s.
\end{align*}
then for each $t\geq0$, the deterministic convolution integral $\int_{0}^{t}S(t-r)u(r)dr$ is well defined $\mathbb{P}$-a.s,  and the deterministic convolution process $\int_{0}^{t}S(t-r)u(r)dr$,~~$t\geq0$,~~is $X$-valued, $\mathbb{F}$-progressively measurable. If for each $t\geq0$, the function $F(t,\cdot):X\rightarrow X$ is bounded, and for each $T>0$ and $x\in X$, the function $F(\cdot,x):[0,T]\rightarrow X$ is Borel measurable, then the process $F(t,u(t))$,~~$t\geq0$,~~is an $X$-valued $\mathbb{F}$-progressively measurable process such that for each $t\geq0$,
\begin{align*}
\int_{0}^{t}\left\|F(r,u(r))\right\|_{X}dr<\infty\quad \mathbb{P}-a.s.
\end{align*}

\noindent Thus, for each $t\geq 0$, the deterministic convolution integral $\int_{0}^{t}S(t-r)F(r,u(r))dr$ is well-defined, and the deterministic convolution process $\int_{0}^{t}S(t-r)F(r,u(r))dr$,~~$t\geq0$,~~is $X$-valued $\mathbb{F}$-progressively measurable.
\\

It follows from \cite{[brzezniak10]} that if $X$ is a Banach space satisfying the $H_{p}$ condition with the norm $\|\cdot\|_{X}$, $\{\tilde{S}(t)\}_{t\geq 0}$ is a contraction $C_{0}$-semigroup on $X$ and $\xi$ is a $\gamma(H,X)$-valued, $\mathbb{F}$-progressively measurable process such that for some $T>0$,
\begin{align*}
\mathbb{E}\int_{0}^{T}\|\xi(r)\|^{2}_{X}dr<\infty,
\end{align*}
then there exists a constant $K>0$ depending on $H$, $X$ and $K_{1}(p),K_{2}(p)$ appearing in the $H_{p}$ condition such that
\begin{align*}
\mathbb{E}\sup_{t\in[0,T]}\left\|\int_{0}^{t}\tilde{S}(t-r)\xi(r)dW(r)\right\|^{2}_{X}\leq K \mathbb{E}\int_{0}^{T}\|\xi(t)\|^{2}_{\gamma(H,X)}dt.
\end{align*}

\noindent Note that A semigroup $\tilde{S}$ defined by $\tilde{S}(t)=e^{-\beta t}S(t)$,~~$t\geq0$,~~is a contraction $C_{0}$-semigroup on $X$. Therefore, we can prove that

\begin{equation}\label{2equa1}
\mathbb{E}\sup_{t\in[0,T]}\left\|\int_{0}^{t}S(t-r)\xi(r)dW(r)\right\|^{2}_{X}\leq K e^{2\beta T}\mathbb{E}\int_{0}^{T}\|\xi(t)\|^{2}_{\gamma(H,X)}dt.
\end{equation}


\section{The existence and the uniqueness of solutions to see (\ref{3equa1}) in Banach spaces with globally Lipschitz coefficients}
\label{s:2.1}

\noindent The main result of this section is the following theorem.

\begin{theorem}\label{3theo1}
Let $X$ be a Banach space satisfying the $H_{p}$ condition endowed with the norm $\|\cdot\|_{X}$. Assume that $\{S(t)\}_{t\geq0}$ is a contraction type $C_{0}$-semigroup on $X$ with the infinitesimal generator $A$. Assume that for each $t\geq0$, $F(t,\cdot)~:~X\rightarrow X$ and $G(t,\cdot)~:~X\rightarrow\gamma(H,X)$ are globally lipschitz maps on $X$, i.e. for each $T>0$ there exist constants $L_{F}>0$ and $L_{G}>0$  such that for all $t\in [0,T]$,
\begin{equation}\label{3equa3}
\|F(t,x_{1})-F(t,x_{2})\|_{X}\leq L_{F}\|x_{1}-x_{2}\|_{X},\quad x_{1},x_{2}\in X,
\end{equation}
\begin{equation}\label{3equa4}
\|G(t,x_{1})-G(t,x_{2})\|_{\gamma(H,X)}\leq L_{G}\|x_{1}-x_{2}\|_{X},\quad x_{1},x_{2}\in X.
\end{equation}

\noindent Moreover, we assume that for every $x\in X$  the functions $F(\cdot,x):[0,\infty)\ni t\mapsto F(t,x)\in X$ and $G(\cdot,x):[0,\infty)\ni t\mapsto G(t,x)\in\gamma(H,X)$ are Borel measurable, and for all  $T>0$ and $x\in X$,
\begin{equation}\label{eqquuaa}
\sup_{t\in[0,T]}\left(\|F(t,x)\|_{X}+\|G(t,x)\|_{\gamma(H,X)}\right)<\infty.
\end{equation}

\noindent Then for each  $u_{0}\in L^{2}(\Omega,\mathcal{F}_{0},\mathbb{P};X)$, there exists a unique $X$-valued mild solution to SEE (\ref{3equa1}).
\end{theorem}

\begin{proof}\noindent \textbf{Existence :}
\noindent The proof for the case $\beta<0$ is included in the proof for the case $\beta\geq0$, since for $\beta<0$,~~$e^{\beta t}\leq1$. Therefore,  we assume that $\{S(t)\}_{t\geq0}$ is a contraction type $C_{0}$-semigroup on $X$, i.e. there exists a constant $\beta\geq0$ such that
\begin{equation}\label{semigroup}
\|S(t)\|_{\mathcal{L}(X)}\leq e^{\beta t},\quad t\geq0.
\end{equation}
By the definition of the mild solution, it is sufficient to show that  equation (\ref{3equa2}) has a unique solution.
\\

For each $T>0$, let $Z_{T}$ be the space of all $X$-valued, $\mathbb{F}$-progressively measurable processes $u$ on $[0,T]$ such that the trajectories of $u$ are $\mathbb{P}$-a.s continuous and
\begin{align*}
\mathbb{E}\sup_{t\in[0,T]}\|u(t)\|^{2}_{X}<\infty.
\end{align*}

Obviously $Z_{T}$ is a Banach space endowed with the norm
\begin{align*}
||u||_{T}=\left(\mathbb{E}\sup_{t\in[0,T]}\|u(t)\|^{2}_{X}\right)^{\frac{1}{2}}.
\end{align*}

\noindent Fix $u_{0}\in L^{2}(\Omega,\mathcal{F}_{0},\mathbb{P};X)$ and $T>0$. Define a map $\Phi_{T,u_{0}}$ by

\begin{align*}
[\Phi_{T,u_{0}}](u)(t)=S(t)u_{0}+\int_{0}^{t}S(t-r)F(r,u(r))dr+\int_{0}^{t}S(t-r)G(r,u(r))dW(r),\quad t\in[0,T],\quad u\in Z_{T}.
\end{align*}

\noindent It follows from \cite{[ball]} and \cite{[brzezniak11]} that the deterministic convolution process $\int_{0}^{t}S(t-r)F(r,u(r))dr$,~~$t\geq0$,~~and the stochastic convolution process $\int_{0}^{t}S(t-r)G(r,u(r))dW(r)$,~~$t\geq0$,~~are well-defined and belong to $Z_{T}$. Moreover, the process $S(\cdot)u_{0}$ belongs to $Z_{T}$ since $\mathbb{E}\|u_{0}\|_{X}^{2}<\infty$. Thus,  $\Phi_{T,u_{0}}$ is a well-defined map from $Z_{T}$ into $Z_{T}$.
Furthermore, we shall prove that the map $\Phi_{T,u_{0}}$ is globally Lipschitz on $Z_{T}$. For this aim let us fix $u_{1},u_{2}\in Z_{T}$. Then, using the Cauchy-Schwarz inequality, (\ref{3equa3}) and (\ref{semigroup})~~we have, for each $t\geq0$,

\begin{align*}
\left\|\int_{0}^{t}S(t-r)[F(r,u_{1}(r))-F(r,u_{2}(r))]dr\right\|_{X}&\leq\int_{0}^{t}\big\|S(t-r)[F(r,u_{1}(r))-F(r,u_{2}(r))]\big\|_{X}dr\\
&\leq\sqrt{t}\left(\int_{0}^{t}e^{2\beta(t-r)}\|F(r,u_{1}(r))-F(r,u_{2}(r))\|^{2}_{X}dr\right)^{\frac{1}{2}}\\
&\leq L_{F}\sqrt{t}e^{\beta t}\left(\int_{0}^{t}\|u_{1}(r)-u_{2}(r)\|^{2}_{X}dr\right)^{\frac{1}{2}}.
\end{align*}

Therefore, we infer that
\begin{equation}\label{eqqqqq1}
\left\|\int_{0}^{t}S(t-r)[F(r,u_{1}(r))-F(r,u_{2}(r))]dr\right\|^{2}_{X}\leq L^{2}_{F}te^{2\beta t}\int_{0}^{t}\|u_{1}(r)-u_{2}(r)\|^{2}_{X}dr,\quad t\geq0
\end{equation}
and thus
\begin{equation}\label{3equa10}
\mathbb{E}\sup_{t\in[0,T]}\left\|\int_{0}^{t}S(t-r)[F(r,u_{1}(r))-F(r,u_{2}(r))]dr\right\|^{2}_{X}\leq L^{2}_{F}T^{2}e^{2\beta T}\|u_{1}-u_{2}\|^{2}_{T}.
\end{equation}

Moreover, using (\ref{2equa1}) and (\ref{3equa4})~~we have

\begin{equation}\label{3equa11}
\mathbb{E}\sup_{t\in[0,T]}\left\|\int_{0}^{t}S(t-r)[G(r,u_{1}(r))-G(r,u_{2}(r))]dW(r)\right\|^{2}_{X}\leq KTL_{G}^{2}e^{2\beta T}\|u_{1}-u_{2}\|^{2}_{T}.
\end{equation}

Taking into account inequalities (\ref{3equa10}) and (\ref{3equa11})~~we infer that
\begin{equation}\label{3equa12}
\|\Phi_{T,u_{0}}(u_{1})-\Phi_{T,u_{0}}(u_{2})\|_{T}\leq C(T)\|u_{1}-u_{2}\|_{T},
\end{equation}
where $C(T)=e^{\beta T}(2L^{2}_{F}T^{2}+2KTL^{2}_{G})^{\frac{1}{2}}$. Thus, $\Phi_{T,u_{0}}$ is globally Lipschitz on $Z_{T}$. If we choose $T$ small enough, say $T_{0}$, such that  $C(T_{0})\leq\frac{1}{2}$, then by the Banach Fixed Point Theorem, there exists a unique process $u^{1}\in Z_{T_{0}}$ such that  $\Phi_{T,u_{0}}(u^{1})=u^{1}$. Therefore, equation (\ref{3equa2}) has a unique solution $u^{1}$ on $[0,T_{0}]$.

Let $Z_{(k-1)T_{0},kT_{0}}$,~~$k=1,2,3...$,~~be the space of all $X$-valued, $\mathbb{F}$-progressively measurable stochastic processes $u$ on $[(k-1)T_{0},kT_{0}]$ such that trajectories of $u$ are $\mathbb{P}$-a.s. continuous and
\begin{align*}
\mathbb{E}\sup_{t\in[(k-1)T_{0},kT_{0}]}\|u(t)\|^{2}_{X}<\infty.
\end{align*}

\noindent For each $k$,~~$Z_{(k-1)T_{0},kT_{0}}$ is a Banach space endowed with the norm
\begin{align*}
\|u\|_{(k-1)T_{0},kT_{0}}=\left(\mathbb{E}\sup_{t\in[(k-1)T_{0},kT_{0}]}\|u(t)\|^{2}_{X}\right)^{\frac{1}{2}}.
\end{align*}

\noindent As shown above, it can be easily shown that the following equation
\begin{equation}\label{3equa13}
u(t)=S(t-(k-1)T_{0})u((k-1)T_{0})+\int_{(k-1)T_{0}}^{t}S(t-r)F(r,u(r))dr+\int_{(k-1)T_{0}}^{t}S(t-r)G(r,u(r))dW(r)
\end{equation}
has a unique solution $u^{k}$ in the space $Z_{(k-1)T_{0},kT_{0}}$ such that $u^{k}(kT_{0})=u^{k+1}(kT_{0})$. Consequently we have a sequence $(u^{k})_{k\in\mathbb{N}}$ of solutions.
\\

Define a process $u$ by

\begin{equation}\label{3equa14}
u(t)=\sum_{k=1}^{\infty}u^{k}(t)1_{[(k-1)T_{0},kT_{0}]}(t),\quad t\in[0,\infty).
\end{equation}

We claim that this process $u$ is a unique solution to equation (\ref{3equa2}).

\begin{proof}
We have already proved that for $k=1$, the process $u$ on $[0,T_{0}]$ solves equation (\ref{3equa2}). By induction, we assume that the process $u$ on $[0,kT_{0}]$ solves equation (\ref{3equa2}) and we will show that the process $u$ on $[0,(k+1)T_{0}]$ solves equation (\ref{3equa2}). We know that $u^{k+1}$ on $[kT_{0},(k+1)T_{0}]$ solves equation (\ref{3equa13}) and $u^{k}(kT_{0})=u^{k+1}(kT_{0})$. Thus, for each $t\in [kT_{0},(k+1)T_{0}]$,~~we obtain

\begin{align*}
u(t)=u^{k+1}(t)=S(t-kT_{0})&\left(S(kT_{0})u_{0}+\int_{0}^{kT_{0}}S(kT_{0}-r)F\left(r,u^{k}(r)\right)dr+\int_{0}^{kT_{0}}S(kT_{0}-r)G\left(r,u^{k}(r)\right)dW(r)\right)\\
&+\int_{kT_{0}}^{t}S(t-r)F\left(t,u^{k+1}(r)\right)dr+\int_{kT_{0}}^{t}S(t-r)G\left(t,u^{k+1}(r)\right)dW(r)\\
&=S(t)u_{0}+\int_{0}^{kT_{0}}S(t-r)F\left(r,u^{k}(r)\right)dr+\int_{0}^{kT_{0}}S(t-r)G\left(r,u^{k}(r)\right)dW(r)\\
&+\int_{kT_{0}}^{t}S(t-r)F\left(t,u^{k+1}(r)\right)dr+\int_{kT_{0}}^{t}S(t-r)G\left(t,u^{k+1}(r)\right)dW(r)\\
&=S(t)u_{0}+\int_{0}^{t}S(t-r)F\left(r,u(r)\right)dr+\int_{0}^{t}S(t-r)G\left(r,u(r)\right)dW(r).
\end{align*}

Therefore, $u$ defined in (\ref{3equa14}) is a solution to equation (\ref{3equa2}).
\\

\textbf{Uniqueness :} In principle, the uniqueness of solutions follows from our proof via the Banach Fixed Point Theorem. However, for completeness and educational purposes, we will present now our independent proof.
\\

Let $u_{1}$ and $u_{2}$ be two solutions to equation (\ref{3equa2}). Define a process $z$ by
\begin{align*}
z(t)=u_{1}(t)-u_{2}(t),\quad t\geq0.
\end{align*}

\noindent Then for each $t\in[0,\infty)$,~~we have
\begin{align*}
&\mathbb{E}\|z(t)\|^{2}_{X}=\mathbb{E}\left\|\int_{0}^{t}S(t-r)[F(t,u_{1}(r))-F(t,u_{2}(r))]dr+\int_{0}^{t}S(t-r)[G(t,u_{1}(r))-G(t,u_{2}(r))]dW(r)\right\|^{2}_{X}\\
&\leq 2\mathbb{E}\left\|\int_{0}^{t}S(t-r)[F(t,u_{1}(r))-F(t,u_{2}(r))]dr\right\|_{X}^{2}+2\mathbb{E}\left\|\int_{0}^{t}S(t-r)[G(t,u_{1}(r))-G(t,u_{2}(r))]dW(r)\right\|^{2}_{X}.
\end{align*}

\noindent By (\ref{equuu1}), (\ref{3equa4}) and (\ref{semigroup}),~~we have,~~for each $t\geq0$,
\begin{align}\label{equuaa}
\begin{split}
\mathbb{E}\left\|\int_{0}^{t}S(t-r)[G(t,u_{1}(r))-G(t,u_{2}(r))]dW(r)\right\|^{2}_{X}&\leq C\mathbb{E}\int_{0}^{t}\|S(t-r)[G(t,u_{1}(r))-G(t,u_{2}(r))]\|^{2}_{\gamma(H,X)}dr\\
&\leq C\mathbb{E}\int_{0}^{t}e^{2\beta(t-r)}\|G(t,u_{1}(r))-G(t,u_{2}(r))\|^{2}_{X}dr\\
&\leq CL_{G}^{2}e^{2\beta t}\int_{0}^{t}\mathbb{E}\|u_{1}(r)-u_{2}(r)\|^{2}_{X}dr.
\end{split}
\end{align}

\noindent It follows from inequalities  (\ref{eqqqqq1}) and (\ref{equuaa}) that

\begin{align*}
\mathbb{E}\|z(t)\|^{2}_{X}\leq K(t)\int_{0}^{t}\mathbb{E}\|z(r)\|^{2}_{X}dr,\quad t\geq0,
\end{align*}

\noindent where $K(t)=2e^{2\beta t}(tL_{F}^{2}+C L_{G}^{2})$. Applying Gronwall's inequality, see \cite{[Ye]},  to function $\mathbb{E}\|z(t)\|^{2}_{X}$,~~$t\geq0$,~~we obtain that for all $t\geq0$,~~~$\mathbb{E}\|z(t)\|^{2}_{X}=0$, which completes the proof of the claim.
\end{proof}

Therefore,  $u$ defined in (\ref{3equa14}) is a unique mild solution to SEE (\ref{3equa1}). Hence the proof of Theorem \ref{3theo1} is complete.
\end{proof}

\section{The existence and the uniqueness of solutions to see (\ref{3equa1}) in Banach spaces with locally Lipschitz coefficients}
\label{s:2.2}
The main result of this section is the following theorem.

\begin{theorem}\label{3theo3}
Let $X$ be a Banach space satisfying the $H_{p}$ condition endowed with the norm $\|\cdot\|_{X}$. Assume that $\{S(t)\}_{t\geq0}$ is a contraction type $C_{0}$-semigroup on $X$ with the infinitesimal generator $A$. Assume that for each $t\geq0$, $F(t,\cdot):X\rightarrow X$ and $G(t,\cdot):X\rightarrow\gamma(H,X)$ are locally Lipschitz maps, i.e. for all  $T>0$ and $R>0$ there exist constants $L_{F}(R)>0$ and $L_{G}(R)>0$ such that if $t\in [0,T]$, $\|x_{1}\|_{X},\|x_{2}\|_{X}\leq R$, then
\begin{equation}\label{3equa15}
\|F(t,x_{1})-F(t,x_{2})\|_{X}\leq L_{F}(R)\|x_{1}-x_{2}\|_{X},
\end{equation}
and
\begin{equation}\label{3equa16}
\|G(t,x_{1})-G(t,x_{2})\|_{\gamma(H,X)}\leq L_{G}(R)\|x_{1}-x_{2}\|_{X}.
\end{equation}

\noindent Moreover, we assume that for each $x\in X$, the functions $F(\cdot,x):[0,\infty)\ni t\rightarrow F(t,x)\in X$ and $G(\cdot,x):[0,\infty)\ni t\rightarrow G(t,x)\in\gamma(H,X)$ are Borel measurable, and for each $t\geq0$, $F(t,\cdot)$ and $G(t,\cdot)$ are of linear growth (uniformly in $t$), that is, for all  $T>0$, there exist constants $\bar{L}_{F}>0$ and $\bar{L}_{G}>0$ such that for all $t\in[0,T]$,
\begin{equation}\label{3equa5}
\|F(t,x)\|^{2}_{X}\leq \bar{L}^{2}_{F}\left(1+\|x\|^{2}_{X}\right),\quad x\in X,
\end{equation}
\begin{equation}\label{3equa6}
\|G(t,x)\|^{2}_{\gamma(H,X)}\leq \bar{L}^{2}_{G}\left(1+\|x\|^{2}_{X}\right),\quad x\in X.
\end{equation}
\noindent Then for each $u_{0}\in L^{2}(\Omega,\mathcal{F}_{0},\mathbb{P};X)$, there exists a unique $X$-valued mild solution $u$ to SEE (\ref{3equa1}).
\end{theorem}

\begin{proof}

\noindent We assume $\beta\geq0$ as in the proof of Theorem \ref{3theo1}. For each $n\in\mathbb{N}$ and $t\geq0$, define mappings $F^{n}(t,\cdot):X\rightarrow X$ and $G^{n}(t,\cdot):X\rightarrow\gamma(H,X)$ by respectively,

\begin{equation}\label{3equa17}
F^{n}(t,x)=
\begin{cases}
F(t,x),\quad \|x\|_{X}\leq n,\\
F\left(t,n\frac{x}{\|x\|_{X}}\right),\quad \|x\|_{X}>n
\end{cases}
\end{equation}
and
\begin{equation}\label{3equa18}
G^{n}(t,x)=
\begin{cases}
G(t,x),\quad\|x\|_{X}\leq n,\\
G\left(t,n\frac{x}{\|x\|_{X}}\right),\quad\|x\|_{X}>n.
\end{cases}
\end{equation}

\noindent For each $n\in\mathbb{N}$ and $t\geq0$, $F^{n}(t,\cdot)$ and $G^{n}(t,\cdot)$ are globally Lipschitz on $X$ with constants $3L_{F}(n)$ and $3L_{G}(n)$ (independent of $t$), see \cite{[brzezniak6]}. Moreover, for each $n\in\mathbb{N}$ and $t\geq0$, $F^{n}(t,\cdot)$ and $G^{n}(t,\cdot)$ satisfy the linear growth conditions (\ref{3equa5}) and (\ref{3equa6}) with the same constants $\bar{L}_{F}$ and $\bar{L}_{G}$. We will  prove linear growth of $F^{n}(t,\cdot)$.
\\

Fix $n\in\mathbb{N}$ and $t\geq0$. Then
\\

\noindent Case 1 : Consider $x\in X$ such that $\|x\|_{X}\leq n$. Then, $F^{n}(t,x)=F(t,x)$. Therefore, from (\ref{3equa5})~~we get
\begin{align*}
\|F^{n}(t,x)\|^{2}_{X}=\|F(t,x)\|^{2}_{X}\leq \bar{L}^{2}_{F}\left(1+\|x\|^{2}_{X}\right).
\end{align*}

\noindent Case 2 : Consider $x\in X$ such that $\|x\|_{X}>n$. Then, $F^{n}(t,x)=F\left(t,n\frac{x}{\|x\|_{X}}\right)$. Therefore, from (\ref{3equa5})~~we get

\begin{align*}
\|F^{n}(t,x)\|^{2}_{X}\leq \bar{L}^{2}_{F}\left(1+n^{2}\left\|\frac{x}{\|x\|_{X}}\right\|^{2}_{X}\right)\leq\bar{L}^{2}_{F}\left(1+\|x\|^{2}_{X}\right).
\end{align*}
Hence, we infer that
\begin{equation}\label{3equa19}
\|F^{n}(t,x)\|^{2}_{X}\leq\bar{L}^{2}_{F}\left(1+\|x\|^{2}_{X}\right),\quad x\in X.
\end{equation}

\noindent Similarly, we can prove that
\begin{equation}\label{3equa20}
\|G^{n}(t,x)\|^{2}_{\gamma(H,X)}\leq \bar{L}^{2}_{G}\left(1+\|x\|^{2}_{X}\right),\quad x\in X.
\end{equation}

For each $n\in\mathbb{N}$, consider the following stochastic evolution equation

\begin{equation}\label{3equa21}
\begin{cases}
du^{n}(t)=[Au^{n}(t)+F^{n}(t,u^{n}(t))]dt+G^{n}(t,u^{n}(t))dW(t),\quad t\geq 0,\\
u^{n}(0)=u_{0}.
\end{cases}
\end{equation}

\noindent By Theorem \ref{3theo1}, SEE (\ref{3equa21}) has a unique $X$-valued mild solution $u^{n}$. Moreover, we shall prove that for each $T>0$, there exists a constant $C(T)>0$ (independent of $n$) such that
\begin{equation}\label{3equa22}
\mathbb{E}\sup_{t\in[0,T]}\|u^{n}(t)\|^{2}_{X}\leq C(T), \quad  n\in\mathbb{N}.
\end{equation}

\noindent \textbf{Proof of (\ref{3equa22}):} Fix $n\in\mathbb{N}$ and $T>0$. Let $u^{n}$ be the unique solution of equation (\ref{3equa21}). Then
\begin{align*}
u^{n}(t)=S(t)u_{0}+\int_{0}^{t}S(t-r)F^{n}(r,u^{n}(r))dr+\int_{0}^{t}S(t-r)G^{n}(r,u^{n}(r))dW(r),\quad t\in[0,\infty).
\end{align*}
Thus, we have, for each $s\in[0,T]$,
\begin{align*}
\mathbb{E}\sup_{t\in[0,s]}\big\|u^{n}(t)\big\|^{2}_{X}&\leq3\mathbb{E}\sup_{t\in[0,s]}\|S(t)u_{0}\|^{2}_{X}+3\mathbb{E}\sup_{t\in[0,s]}\left\|\int_{0}^{t}S(t-r)F^{n}(r,u^{n}(r))dr\right\|^{2}_{X}\\
&+3\mathbb{E}\sup_{t\in[0,s]}\left\|\int_{0}^{t}S(t-r)G^{n}(r,u^{n}(r))dW(r)\right\|^{2}_{X}.
\end{align*}

\noindent By the Cauchy-Schwarz inequality, (\ref{semigroup}) and (\ref{3equa19}),~~we obtain, for each $t\geq0$,
\begin{align*}
\left\|\int_{0}^{t}S(t-r)F^{n}\left(r,u^{n}(r)\right)dr\right\|_{X}&\leq\int_{0}^{t}\big\|S(t-r)F^{n}\left(r,u^{n}(r)\right)\big\|_{X}dr\\
&\leq\bar{L}_{F}\sqrt{t}e^{\beta t}\left(\int_{0}^{t}\left(1+\big\|u^{n}(r)\big\|^{2}_{X}\right)dr\right)^{\frac{1}{2}}\\
&\leq t\bar{L}_{F}e^{\beta t}+\bar{L}_{F}\sqrt{t}e^{\beta t}\left(\int_{0}^{t}\big\|u^{n}(r)\big\|^{2}_{X}dr\right)^{\frac{1}{2}}\\
&\leq t\bar{L}_{F}e^{\beta t}+\bar{L}_{F}\sqrt{t}e^{\beta t}\left(\int_{0}^{t}\sup_{t\in[0,r]}\big\|u^{n}(t))\big\|^{2}_{X}dr\right)^{\frac{1}{2}}.
\end{align*}
Therefore, we obtain
\begin{equation}\label{eqqq1}
\mathbb{E}\sup_{t\in[0,s]}\left\|\int_{0}^{t}S(t-r)F^{n}\left(r,u^{n}(r)\right)dr\right\|^{2}_{X}\leq 2s^{2}\bar{L}^{2}_{F}e^{2\beta s}+2s\bar{L}^{2}_{F}e^{2\beta s}\int_{0}^{s}\mathbb{E}\sup_{t\in[0,r]}\big\|u^{n}(t))\big\|^{2}_{X}dr.
\end{equation}

\noindent By (\ref{2equa1}),~~(\ref{semigroup})~~and (\ref{3equa20}),~~we have
\begin{align}\label{eqqq2}
\begin{split}
\mathbb{E}\sup_{t\in[0,s]}\left\|\int_{0}^{t}S(t-r)G^{n}\left(r,u^{n}(r)\right)dW(r)\right\|^{2}_{X}&\leq Ke^{2\beta s}\mathbb{E}\int_{0}^{s}\big\|G^{n}\left(r,u^{n}(r)\right)\big\|_{\gamma(H,X)}^{2}dr\\
&\leq K\bar{L}^{2}_{G}e^{2\beta s}\mathbb{E}\int_{0}^{s}\left(1+\big\|u^{n}(r)\big\|^{2}_{X}\right)dr\\
&\leq sK\bar{L}^{2}_{G}e^{2\beta s}+K\bar{L}^{2}_{G}e^{2\beta s}\int_{0}^{s}\mathbb{E}\sup_{t\in[0,r]}\big\|u^{n}(t))\big\|^{2}_{X}dr.
\end{split}
\end{align}

\noindent It follows from inequalities (\ref{eqqq1}) and (\ref{eqqq2}) that
\begin{align*}
\mathbb{E}\sup_{t\in[0,s]}\big\|u^{n}(t)\big\|^{2}_{X}&\leq M(s)+L(s)\int_{0}^{s}\mathbb{E}\sup_{t\in[0,r]}\big\|u^{n}(t)\big\|^{2}_{X}dr,\quad s\in[0,T],
\end{align*}
where $M(s)=3e^{2\beta s}\left(\mathbb{E}\|u_{0}\|^{2}_{X}+2s^{2}\bar{L}^{2}_{F}+sK\bar{L}^{2}_{G}\right)$ and $L(s)=3e^{2\beta s}\left(2\bar{L}^{2}_{F}s+K\bar{L}^{2}_{G}\right)$.~~Applying  Gronwall's inequality~~we infer that
\begin{align*}
\mathbb{E}\sup_{t\in[0,s]}\big\|u^{n}(t)\big\|^{2}_{X}\leq M(s)e^{L(s)s},\quad s\in[0,T].
\end{align*}
This gives the desired conclusion.
\\

For each $n\in\mathbb{N}$, define a map $\tau_{n} : \Omega\rightarrow [0,\infty]$ by

\begin{align*}
\tau_{n}(\omega)=\inf\left\{t\in[0,\infty)~:~\left\|u^{n}(t,\omega)\right\|_{X}\geq n\right\},\quad \omega\in\Omega.
\end{align*}

\noindent For each $n\in\mathbb{N}$, $\tau_{n}$ is a stopping time, see \cite{[Karatzas]} for the proof. Moreover, we shall prove that the sequence $(\tau_{n})_{n\in\mathbb{N}}$ converges to $\infty$. For this aim it is sufficient to show that for each $T>0$,

\begin{equation}\label{3equa26}
\mathbb{P}\left(\left\{\omega\in\Omega~|~\exists~k\in\mathbb{N} : \forall~n\geq k~~\tau_{n}(\omega)\geq T\right\}\right)=1.
\end{equation}

\noindent Fix $T>0$. Then using inequality (\ref{3equa22}) and the Chebyshev Inequality~~we have

\begin{align*}
\mathbb{P}\left(\left\{\omega\in\Omega : \sup_{t\in[0,T]}\|u^{n}(t)\|_{X}\geq n\right\}\right)\leq C(T)\frac{1}{n^{2}}.
\end{align*}

\noindent Set
\begin{align*}
A_{n}=\left\{\omega\in\Omega : \sup_{t\in[0,T]}\|u^{n}(t)\|_{X}\geq n\right\}.
\end{align*}

\noindent Since $\sum_{n=1}^{\infty}\frac{1}{n^{2}}<\infty$, $\sum_{n=1}^{\infty}\mathbb{P}(A_{n})<\infty$. Therefore,  by the Borel Cantelli Lemma, $\mathbb{P}\left(\cap_{k=1}^{\infty}\cup_{n=k}^{\infty}A_{n}\right)=0$. Thus $\mathbb{P}\left(\cup_{k=1}^{\infty}\cap_{n=k}^{\infty}(\Omega\setminus A_{n})\right)=1$. Take $\omega\in \cup_{k=1}^{\infty}\cap_{n=k}^{\infty}(\Omega\setminus A_{n})$. Then $\exists~k\in\mathbb{N}$ such that  $\omega\in\cap_{n=k}^{\infty}(\Omega\setminus A_{n})$, i.e. $\forall~n\geq k$,~~$\omega\in\Omega\setminus A_{n}$. Therefore, $\forall~n\geq k$, $\sup_{t\in[0,T]}\left\|u^{n}(t,\omega)\right\|_{X}< n$. Thus, $\exists~k\in\mathbb{N}$ such that $\forall~n\geq k$, $\left\|u^{n}(t,\omega)\right\|_{X}< n$,~~$t\in[0,T]$. Therefore, $\exists~k\in\mathbb{N}$ such that $\forall~n\geq k$, $\tau_{n}(\omega)> T$. This gives the desired conclusion.
\\

Define a process $u$ by

\begin{equation}\label{3equa27}
u(t)=u^{n}(t),\quad if~~t\leq\tau_{n}.
\end{equation}

\noindent Note that by \cite{[brzezniak7]}, $u^{n}$ solves the following equation
\begin{align*}
u(t\wedge \tau_{n})=u^{n}(t\wedge \tau_{n})&=S(t\wedge \tau_{n})u_{0}+\int_{0}^{t\wedge \tau_{n}}S(t\wedge \tau_{n}-r)F^{n}\left(r,u^{n}(r)\right)dr\\
&+\int_{0}^{t\wedge \tau_{n}}S(t\wedge \tau_{n}-r)G^{n}\left(r,u^{n}(r)\right)dW(r),\quad t\geq0.
\end{align*}

\noindent Since $r\leq \tau_{n}$ and by the definition of $\tau_{n}$,~~$\big\|u^{n}(r)\big\|_{X}\leq n$. Therefore, from the definition of $F^{n}(t,\cdot)$ and $G^{n}(t,\cdot)$~~we get

\begin{align*}
F^{n}\left(r,u^{n}(r)\right)=F(r,u^{n}(r))~~~~and~~~~G^{n}\left(r,u^{n}(r)\right)=G(r,u^{n}(r)).
\end{align*}

\noindent Also by the definition of $u$,~~$u(r)=u^{n}(r)$ if $r\leq \tau_{n}$. Hence, we obtain
\begin{align*}
u(t\wedge \tau_{n})&=S(t\wedge \tau_{n})u_{0}+\int_{0}^{t\wedge \tau_{n}}S(t\wedge \tau_{n}-r)F(r,u(r))dr+\int_{0}^{t\wedge \tau_{n}}S(t\wedge \tau_{n}-r)G(r,u(r))dW(r).
\end{align*}

\noindent Since $\tau_{n}\rightarrow\infty$,~~$t\wedge \tau_{n}\rightarrow t$. Thus,~~$u(t\wedge \tau_{n})\rightarrow u(t)$ and $S(t\wedge \tau_{n})\rightarrow S(t)$~~as~$n\rightarrow\infty$. Therefore, we infer that
\begin{align*}
u(t)=S(t)u_{0}+\int_{0}^{t}S(t-r)F(r,u(r))dr+\int_{0}^{t}S(t-r)G(r,u(r))dW(r).
\end{align*}

\noindent Thus $u$ defined in (\ref{3equa27}) is a mild solution to SEE (\ref{3equa1}). The uniqueness of $u$ follows from Theorem \ref{3theo1}.
\end{proof}

\section{Markov property and ergodic properties of  the solutions to see (\ref{3equa1})}
\label{sec:Markov}
For $x\in X$ and $0\leq s<\infty$, we denote by $u(t,s;x)$, $t\geq s$ the unique solution of SEE (\ref{3equa1}) on the time interval $[s,\infty)$ with the initial date $u(s)=x$. Let $C_{b}(X)$ be the set of all bounded measurable functions $\varphi$ from $X$ to $\mathbb{R}$.

\begin{definition}
A family of  maps $P_{s,t}:C_{b}(X)\rightarrow \mathbb{R}$, $ t\geq s\geq0$, defined by
\begin{align*}
P_{s,t}(\varphi)(x)=\mathbb{E}\bigl[\varphi(u(t,s;x))\bigr],\quad x\in X,\quad\varphi\in C_{b}(X)
\end{align*}
is called the transition semigroup corresponding to SEE (\ref{3equa1}).
\end{definition}

\begin{definition}
An $X$-valued process $u$ is called Markov if for each $\varphi\in C_{b}(X)$ and $0\leq s\leq t$,
\begin{align*}
\mathbb{E}(\varphi(u(t))|\mathcal{F}_{s})=P_{t,s}\varphi(u(s))\quad \mathbb{P}-a.s.
\end{align*}
\end{definition}

Similar to Theorem 9.8 in \cite{[prato]}, we have the following result, see \cite{[Kok]} for the proof.

\begin{theorem}\label{theoremm1}
Assume that all the assumptions of Theorem \ref{3theo1} or Theorem \ref{3theo3} are satisfied. Let $u$ be the unique solution of SEE (\ref{3equa1}). \dela{For $x\in X$ and $0\leq r\leq t<\infty$, we denote by $u(t,r;u_{r})$ the value at time $t$ of the solution $u$ with $u(r)=u_{r}\in L^{2}(\Omega,\mathcal{F}_{r},\mathbb{P};X)$.} Then for arbitrary $\varphi\in C_{b}(X)$, $r\geq 0$ and  $u_{r}\in L^{2}(\Omega,\mathcal{F}_{r},\mathbb{P};X)$,
\begin{align*}
\mathbb{E}\left[\varphi(u(t,r;u_{r}))|\mathcal{F}_{s}\right]=P_{s,t}\varphi(u(s,r;u_{r})),\quad \mathbb{P}-a.s. \mbox{ for all} ~~t \geq s\geq r.
\end{align*}
\end{theorem}

\begin{corollary}
Under the all assumptions of Theorem \ref{3theo1} or Theorem \ref{3theo3},~~the solution of SEE (\ref{3equa1}) is  a Markov process.
\end{corollary}

\subsection{Existence and uniqueness of an invariant measure to SEE (\ref{3equa1})}
In this section we consider the following SEE
\begin{equation}\label{4equa}
\begin{cases}
du=(Au+F(u))dt+G(u)dW(t),\quad t\geq 0,\\
u(0)=u_{0},
\end{cases}
\end{equation}
where $F$ and $G$ are time independent. It follows from Theorem \ref{3theo1} or Theorem \ref{3theo3} that if the functions $F:X\rightarrow X$ and $G:X\rightarrow\gamma(H,X)$ satisfy all the assumptions of Theorem \ref{3theo1} or Theorem \ref{3theo3}, then SEE (\ref{4equa}) has a unique $X$-valued mild solution.

\begin{definition}
 Let $P_{t}$, $t\geq 0$, be the translation semigroup corresponding to SEE (\ref{4equa}). A Borel probability measure $\mu$ on $X$ is called an invariant measure for SEE (\ref{4equa}) iff
\begin{align*}
P_{t}^{*}\mu=\mu,\quad t\geq0,
\end{align*}
where $P_{t}^{*}$ is the dual semigroup defined on the space M(X) of all bounded measures on $(X,\mathcal{B}(X))$, and is defined by
\begin{align*}
\langle \varphi,P_{t}^{*}\mu\rangle=\langle P_{t}\varphi,\mu\rangle,\quad \varphi\in C_{b}(X),\quad \mu\in M(X).
\end{align*}
\end{definition}

The following theorem obtained  in  \cite{[brzezniak8]}  gives a sufficient condition for the existence and the uniqueness of invariant measures for the solution to problem (\ref{4equa}),~~see Theorem 3.7 and Theorem 4.6 therein for the proof.

\begin{theorem}\label{theoremm2}
 Assume that the functions $F:X\rightarrow X$ and $G:X\rightarrow\gamma(H,X)$, satisfy all the assumptions of Theorem \ref{3theo1} or Theorem \ref{3theo3}. If there exist a constant $\omega>0$ and a natural number $n_{0}\in\mathbb{N}$ such that for all $x_{1},x_{2}\in X$ and $n\geq n_{0}$,
\begin{equation}\label{equu5}
[A_{n}(x_{1}-x_{2})+F(x_{1})-F(x_{2}),x_{1}-x_{2}]+\frac{K_{2}(p)}{p}
\|G(x_{1})-G(x_{2})\|^{2}_{\gamma(H,X)}\leq-\omega\|x_{1}-x_{2}\|_{\nu,p}^{2},
\end{equation}
 where $A_{n}$ is the Yosida approximation of $A$, and $[\cdot,\cdot]$ is the semi-inner product on $X\times X$, see Definition \ref{def-semi inner product} below, then there exists a unique invariant measure for the solution to SEE (\ref{4equa}).
\end{theorem}

\begin{definition}\label{def-semi inner product}
A semi-inner product on a complex or real vector space $V$ is a mapping $[\cdot,\cdot]:V\times V\rightarrow \mathbb{C}$ (or $\mathbb{R}$) such that

\noindent (i) $[x+y,z]=[x,z]+[y,z]$,\quad $x,y,z\in V$,

\noindent (ii) $[\lambda x,y]=\lambda[x,y]$,\quad $x,y\in V$,\quad$\lambda\in \mathbb{C}~~(or~\mathbb{R})$,

\noindent (iii) [x,x]>0,\quad for $x\neq0$,

\noindent (iv) $|[x,y]|^{2}\leq[x,x]~[y,y]$, \quad$x,y\in V$.
\\

\noindent Such a vector space $V$ is called a semi-inner product space.
\end{definition}

\begin{remark}
A semi-inner product on a Banach space $X$ is given by
\begin{align*}
[x,y]=\langle x,y^{*}\rangle,\quad x,y\in X,
\end{align*}
where $y^{*}\in X^{*}$ such that $\left\|y^{*}\right\|=\|y\|_{X}$ and $\langle y,y^{*}\rangle=\|y\|_{X}$. Such a $y^{*}\in X^{*}$ exists by the Hahn-Banach theorem.
\end{remark}

\section{Application to Heath-Jarrow-Morton-Musiela (HJMM) equation}
\label{s:3}

\subsection{The HJMM equation}
\label{s:3.1}
The value of one dollar at time $t\in[0,T]$ with maturity $T \geq 0$ is called the zero-coupon bond, and  is denoted by $P(t,T)$. This is a contract that guarantees the holder one dollar to be paid at the maturity date $T$. It is assumed that for each $T\geq 0$ and $t\in[0,T]$,~~$P(t,T)$ denotes an $\mathbb{R}$-valued random variable defined on a probability space $(\Omega,\mathcal{F},\mathbb{P})$. Therefore, for each $T>0$, $\{P(t,T)\}_{t\in[0,T]}$ denotes an $\mathbb{R}$-valued stochastic process. Under some assumptions, $P(t,T)$ can be written as follows, see \cite{[Filipovic2]},
\begin{align*}
P(t,T)=e^{-\int_{t}^{T}f(t,s)ds}.
\end{align*}
where for each $0\leq t\leq T$,~~$f(t,T)$ is an $\mathbb{R}$-valued random variable on $(\Omega,\mathcal{F},\mathbb{P})$, and called forward rate. Therefore, for each $T\geq0$, the family $\{f(t,T)\}_{t\in[0,T]}$ of forward rates is a stochastic process, and called forward rate process. In the framework of Heath-Jarrow-Morton \cite{[Heath]}, it was assumed that for each $T\geq0$,  forward rate process $f(t,T)$,~~$t\in[0,T]$, satisfies the following stochastic differential equation,

\begin{equation}\label{equation1}
df(t,T)=\left\langle \sigma(t,T),\int_{t}^{T}\sigma(t,v)dv\right\rangle dt+\left\langle\sigma(t,T),dW(t)\right\rangle,\quad t\in[0,T],
\end{equation}
where $W$ is a standard $d$-dimensional Brownian motion and $\langle\cdot,\cdot\rangle$ denotes the usual inner product in $\mathbb{R}^{d}$. Using the Musiela parametrization \cite{[Musiela]} an important connection between HJM model and stochastic partial differential equations can be provided. Define
\begin{align*}
r(t)(x)=f(t,t+x),\quad T=t+x\quad x,t\geq 0,
\end{align*}
where $x$ is called time to maturity, and for each $t\geq0$,~~the function $r(t)$ is a random variable on $(\Omega,\mathcal{F},\mathbb{P})$ taking values in the space of functions of $x$, and called forward curve. Therefore,~~the family $\{r(t)\}_{t\geq0}$ of forward curves is a stochastic process taking values in the space of functions of $x$, and called forward curve process. By the framework of Heath-Jarrow-Morton \cite{[Heath]}, the forward curve process satisfies the following stochastic partial differential equation,
\begin{equation}\label{equation3}
dr(t)(x)=\left(\frac{\partial }{\partial x}r(t)(x)+\left\langle \alpha(t)(x),\int_{0}^{x}\alpha(t)(y)dy\right\rangle_{H}\right)dt+\langle \alpha(t)(x),dW(t)\rangle_{H},\quad t\geq 0,
\end{equation}
where $\alpha(t)(x)=\sigma(t,t+x)$, $t,x\geq0$, $(H,\langle\cdot,\cdot\rangle_{H})$ is a (possibly infinite dimensional) Hilbert space and $W$ is an $\mathbb{F}$-cylindrical canonical Wiener process on $H$, see \cite{[Kok]} for the derivation of equation (\ref{equation3}).
\\

If the volatility $\alpha$ depends on the forward curve $r$, i.e. $\alpha(t)(x)=\zeta(t,r(t))(x)$, then equation (\ref{equation3}) becomes
\begin{equation}\label{equationnnnnnn5}
dr(t)(x)=\left(\frac{\partial }{\partial x}r(t)(x)+\left\langle \zeta(t,r(t))(x),\int_{0}^{x}\zeta(t,r(t))(y)\,dy\right\rangle_{H}\right)dt+\langle\zeta(t,r(t))(x),dW(t)\rangle_{H}.
\end{equation}

\noindent In this paper we will analyse, for a certain function $\zeta$,  the existence and the uniqueness of solutions to equation (\ref{equationnnnnnn5}) in some certain spaces of functions, and the existence and the uniqueness of invariant measures for the solution to equation (\ref{equationnnnnnn5}).

\subsection{Existence and uniqueness of solutions to HJMM equation with globally Lipschitz coefficients}
\label{s:3.2}

For each $\nu\in\mathbb{R}$ and $p\geq 1$, let $L_{\nu}^{p}$ be the space of all (equivalence classes of) Lebesgue measurable functions $f~: ~[0,\infty)\rightarrow\mathbb{R}$ such that
\begin{align*}
\int_{0}^{\infty}|f(x)|^{p}e^{\nu x}dx<\infty.
\end{align*}
It is well know that for each $\nu\in\mathbb{R}$ and $p\geq1$, $L_{\nu}^{p}$ is a Banach space endowed with the norm
\begin{align*}
\|f\|_{\nu,p}=\left(\int_{0}^{\infty}|f(x)|^{p}e^{\nu x}dx\right)^{\frac{1}{p}}.
\end{align*}
If $p\geq2$, then $L_{\nu}^{p}$ satisfies the $H_{p}$ condition \cite{[brzezniak10]}. Moreover, see \cite{[brzezniak12]}, if $\psi:L^{p}_{\nu}\ni f\mapsto \psi(f)=\|f\|_{\nu,p}^{p}\in\mathbb{R}$, then
\begin{equation}\label{eee10}
|\psi^{\prime}(f)|\leq p\|f\|_{\nu,p}^{p-1}~~~and~~~|\psi^{\prime\prime}(f)|\leq p (p-1)\|f\|_{\nu,p}^{p-2},\quad f\in L^{p}_{\nu}.
\end{equation}

\begin{proposition}\label{prr1}
Let $\nu>0$ and $p\geq1$. Then the space $L_{\nu}^{p}$ is continuously embedded into the space $L^{1}([0,\infty))$, and for all $f\in L_{\nu}^{p}$,
\begin{align*}
||f||_{L^{1}}\leq\left(\frac{p}{\nu q}\right)^{\frac{1}{q}}||f||_{\nu,p},
\end{align*}
where $q\in[1,\infty)$ such that $\frac{1}{p}+\frac{1}{q}=1$.
\end{proposition}

\begin{proof}
Fix $\nu>0$, $p\geq1$ and $f\in L_{\nu}^{p}$. Then by H$\ddot{o}$lder inequality, we get
\begin{align*}
\|f\|_{L^{1}}&=\int_{0}^{\infty}|f(x)|dx=\int_{0}^{\infty}|f(x)|e^{\frac{\nu x}{p}}e^{-\frac{\nu x}{p}}dx\\
&\leq\left(\int_{0}^{\infty}|f(x)|^{p}e^{\nu x}dx\right)^{\frac{1}{p}}\left(\int_{0}^{\infty}e^{-\frac{\nu q x}{p}}dx\right)^{\frac{1}{q}}\\
&=\left(\frac{p}{\nu q}\right)^{\frac{1}{q}}\|f\|_{\nu,p}.
\end{align*}
This gives the desired conclusion.
\end{proof}

\begin{lemma}\label{llem2}
Let $S=\{S(t)\}_{t\geq 0}$ be a family of operators on $L_{\nu}^{p}$ defined by

\begin{equation}\label{shift-semigroup}
S(t)f(x)=f(t+x),\quad f\in L_{\nu}^{p},\quad t,x\in[0,\infty).
\end{equation}
Then $S$ is a contraction type $C_{0}$-semigroup on $L_{\nu}^{p}$ such that
\begin{align*}
\|S(t)\|_{\mathcal{L}(L^{p}_{\nu})}\leq e^{\frac{-\nu t}{p}},\quad t\geq0.
\end{align*}
Moreover, the infinitesimal generator $A$ of $S$ on $L_{\nu}^{p}$ is given by
\begin{align*}
D(A)=\left\{f\in L^{p}_{\nu} : Df\in L^{p}_{\nu}\right\},
\end{align*}
\begin{align*}
Af=Df,\quad f\in D(A),
\end{align*}
where $Df$ is the first weak derivative of $f$.
\end{lemma}

\begin{remark}
The semigroup $S$ defined in (\ref{shift-semigroup}) is called the shift semigroup.
\end{remark}
For the suitability of our aim, throughout this section we assume that $p\geq2$ and $\nu>0$.
\\

Assume that function $\zeta$ in equation (\ref{equationnnnnnn5}) is defined by
\begin{align*}
\zeta(t,f)(x)=g(t,x,f(x)),\quad f\in L^{p}_{\nu},\quad t,x\in[0,\infty),
\end{align*}
where $g:[0,\infty)\times[0,\infty)\times\mathbb{R}\rightarrow H$ is a given function. Define

\begin{equation}\label{eee1}
G(t,f)[h](x)=\left\langle g(t,x,f(x)),h\right\rangle_{H},\quad f\in L^{p}_{\nu},\quad h\in H,\quad t,x\in[0,\infty)
\end{equation}
and
\begin{equation}\label{ett}
F(t,f)(x)=\left\langle g(t,x,f(x)), \int_{0}^{x}g(t,y,f(y))dy\right\rangle_{H},\quad f\in L^{p}_{\nu},\quad t,x\in[0,\infty).
\end{equation}
Then the abstract form of equation (\ref{equationnnnnnn5}) in $L^{p}_{\nu}$  can be written as follows
\begin{equation}\label{equationn5}
dr(t)=\left[Ar(t)+F(t,r(t))\right]dt+G(t,r(t))dW(t),\quad t\in[0,\infty).
\end{equation}

\begin{theorem}\label{theo1}
Let $G$ and $F$ be as given in (\ref{eee1}) and (\ref{ett}) respectively. Assume that there exist functions $\bar{g}\in L_{\nu}^{p}$ and $\hat{g}\in L_{\nu}^{p}\cap L^{\infty}$ such that for all $t\in[0,\infty)$,
\begin{equation}\label{eq12}
\|g(t,x,u)\|_{H}\leq|\bar{g}(x)|,\quad u\in\mathbb{R},\quad x\in[0,\infty)
\end{equation}
and
\begin{equation}\label{eq13}
\|g(t,x,u)-g(t,x,\rm{v})\|_{H}\leq|\hat{g}(x)|~|u-\rm{v}|,\quad u,\rm{v}\in\mathbb{R},\quad x\in[0,\infty).
\end{equation}

\noindent Then for each $r_{0}\in L^{2}(\Omega,\mathcal{F}_{0},\mathbb{P};L^{p}_{\nu})$, there exists a unique $L^{p}_{\nu}$-valued mild solution $r$ to equation (\ref{equationn5}) with $r(0)=r_{0}$ \dela{in the spaces $L^{p}_{\nu}$}. Moreover, the solution is a Markov process.
\end{theorem}

In view of Theorem \ref{3theo1}, the proof of the existence and the uniqueness of solutions follows from Lemma \ref{llem2} and the following lemma. The proof of the Markov property follows from Theorem \ref{theoremm1}.

\begin{lemma}\label{llem3}
Under the assumptions of Theorem \ref{theo1}, for each $t\geq0$, $F(t,\cdot) :L_{\nu}^{p}\rightarrow L_{\nu}^{p}$ and $G(t,\cdot) :L_{\nu}^{p}\rightarrow \gamma(H,L_{\nu}^{p})$ are well-defined. Moreover,~~for all $t\geq 0$,
\\

\noindent (i)
\begin{equation}\label{equaa2}
\|F(t,f)\|_{\nu,p}+\|G(t,f)\|_{\gamma(H,L_{\nu}^{p})}\leq \left(\frac{p}{\nu q}\right)^{\frac{1}{q}}\|\bar{g}\|^{2}_{\nu,p}+N \|\bar{g}\|_{\nu,p},\quad f\in L_{\nu}^{p}
\end{equation}

\noindent (ii) $F(t,\cdot)$ and $G(t,\cdot)$ are globally Lipschitz on $L^{p}_{\nu}$ with Lipschitz constants independent of $t\in[0,\infty]$.
\end{lemma}

\begin{proof}
Our proof is based mainly on  the proposition taken from \cite{[brzezniak3]}
\begin{proposition}\label{2pro2}
Let $H$ be a separable Hilbert space and $\mathcal{O}\subset \mathbb{R}^{d}$. Let $(\mathcal{O},\mathcal{F},\mu)$ be a measurable space and $L^{p}=L^{p}(\mathcal{O},\mathcal{F},\mu;\mathbb{R})$, $p\in[2,\infty)$. Then $K~:~H\rightarrow L^{p}$ defined by
\begin{align*}
K[h](x)=\langle\kappa(x),h\rangle_{H},\quad h\in H,\quad x\in\mathcal{O},
\end{align*}
where $\kappa\in L^{p}(\mathcal{O},\mathcal{F},\mu;H)$,~~is a $\gamma$-radonifying operator from $H$ into $L_{\nu}^{p}$  and for some $N>0$,
\begin{align*}
\|K\|_{\gamma(H,L^{p})}\leq N\left(\int_{\mathcal{O}}\|\kappa(x)\|^{p}_{H}d\mu(x)\right)^{\frac{1}{p}}.
\end{align*}
\end{proposition}

\textbf{Proof of (i) :} Fix $t\geq0$ and $f\in L_{\nu}^{p}$. Then using (\ref{eq12}) we get
\begin{align*}
\int_{0}^{\infty}\|g(t,x,f(x)\|_{H}^{p}e^{\nu x}dx\leq\int_{0}^{\infty}|\bar{g}(x)|^{p}e^{\nu x}dx=\|\bar{g}\|^{p}_{\nu,p}.
\end{align*}
Therefore, by Proposition \ref{2pro2},~~$G(t,f)$ is a $\gamma$-radonifying operator from $H$ into $L_{\nu}^{p}$ and for some $N>0$,
\begin{equation}\label{g}
\|G(t,f)\|_{\gamma(H,L_{\nu}^{p})}\leq N \|\bar{g}\|_{\nu,p}.
\end{equation}

\noindent Thus,~~$G(t,\cdot)$ is well defined. Using the Cauchy-Schwarz inequality~~we get
\begin{align*}
|F(t,f)(x)|\leq \|g(t,x,f(x)\|_{H}\int_{0}^{\infty}\left\|g(t,y,f(y))\right\|_{H}dy.
\end{align*}

\noindent It follows from Proposition \ref{prr1} and (\ref{eq12})  that
\begin{align*}
|F(t,f)(x)|\leq |\bar{g}(x)|\int_{0}^{\infty}|\bar{g}(y)|dy\leq \left(\frac{p}{\nu q}\right)^{\frac{1}{q}}\|\bar{g}\|_{\nu,p}|\bar{g}(x)|.
\end{align*}

\noindent Therefore, we obtain

\begin{equation}\label{eeqq9}
\|F(t,f)(x)\|_{\nu,p}\leq \left(\frac{p}{\nu q}\right)^{\frac{1}{q}}\|\bar{g}\|_{\nu,p}\left(\int_{0}^{\infty}|\bar{g}(x)|^{p}e^{\nu x}dx\right)^{\frac{1}{p}}=\left(\frac{p}{\nu q}\right)^{\frac{1}{q}}\|\bar{g}\|^{2}_{\nu,p}.
\end{equation}

\noindent Hence,~~$F(t,\cdot)$ is well-defined,~~and by inequalities (\ref{g}) and (\ref{eeqq9}),~~we infer that
\begin{align*}
\|F(t,f)\|_{\nu,p}+\|G(t,f)\|_{\gamma(H,L_{\nu}^{p})}\leq \left(\frac{p}{\nu q}\right)^{\frac{1}{q}}\|\bar{g}\|^{2}_{\nu,p}+N \|\bar{g}\|_{\nu,p}.
\end{align*}

\textbf{Proof of (ii) :} Fix $t\geq 0$ and $f_{1},f_{2}\in L_{\nu}^{p}$. Then
\begin{align*}
\left(G(t,f_{1})-G(t,f_{2})\right)[h](x)=\langle g(t,x,f_{1}(x))-g(t,x,f_{2}(x)),h\rangle_{H},\quad x\in[0,\infty),\quad h\in H.
\end{align*}

\noindent From (\ref{eq13}) we get
\begin{equation}\label{eeeqqq1}
\int_{0}^{\infty}\|g(t,x,f_{1}(x))-g(t,x,f_{2}(x))\|_{H}^{p}e^{\nu x}dx\leq \int_{0}^{\infty}|\hat{g}(x)|^{p}|f_{1}(x)-f_{2}(x)|^{p}e^{\nu x}dx\leq\|\hat{g}\|^{p}_{L^{\infty}} \|f_{1}-f_{2}\|^{p}_{\nu,p}.
\end{equation}

\noindent Therefore, by Proposition \ref{2pro2}, there exist a constant $N>0$ such that
\begin{equation}\label{ee5}
\|G(t,f_{1})-G(t,f_{2})\|_{\gamma(H,L^{p}_{\nu})}\leq N\|\hat{g}\|_{L^{\infty}} \|f_{1}-f_{2}\|_{\nu,p}.
\end{equation}

\noindent Hence,~~$G(t,\cdot)$ is globally Lipschitz on $L^{p}_{\nu}$.
\\

Using  the Cauchy-Schwarz inequality,~~(\ref{eq12}) and (\ref{eq13})~~we get

\begin{align*}
|F(t,f_{1})(x)-F(t,f_{2})(x)|&=\left|\left\langle g(t,x,f_{1}(x)),\int_{0}^{x}(g(t,y,f_{1}(y))-g(t,y,f_{2}(y)))dy\right\rangle_{H}\right|\\
&+\left|\left\langle g(t,x,f_{1}(x))-g(t,x,f_{2}(x)),\int_{0}^{x}g(t,y,f_{2}(y))dy\right\rangle_{H}\right|\\
&\leq \|g(t,x,f_{1}(x))\|_{H}\int_{0}^{\infty}\|g(t,x,f_{1}(x))-g(t,x,f_{2}(x))\|_{H}dx\\
&+\|g(t,x,f_{1}(x))-g(t,x,f_{2}(x))\|_{H}\int_{0}^{\infty}\|g(t,x,f_{2}(x))\|_{H}dx\\
&\leq \|\hat{g}\|_{L^{\infty}}|\bar{g}(x)|\int_{0}^{\infty}|f_{1}(x)-f_{2}(x)|dx+|\hat{g}(x)|~|f_{1}(x)-f_{2}(x)|\int_{0}^{\infty}|\bar{g}(x)|dx.
\end{align*}

\noindent It follows from Proposition \ref{prr1} that

\begin{equation}
|F(t,f_{1})(x)-F(t,f_{2})(x)|\leq \left(\frac{p}{\nu q}\right)^{\frac{1}{q}}\|\hat{g}\|_{L^{\infty}}\|f_{1}-f_{2}\|_{\nu,p}|\bar{g}(x)|+\left(\frac{p}{\nu q}\right)^{\frac{1}{q}}\|\bar{g}\|_{\nu,p}|\hat{g}(x)|~|f_{1}(x)-f_{2}(x)|.
\end{equation}

\noindent Therefore, we obtain
\begin{align*}
\|F(t,f_{1})-F(t,f_{2})\|_{\nu,p}&\leq 2\left(\frac{p}{\nu q}\right)^{\frac{1}{q}}\|\hat{g}\|_{L^{\infty}}\|f_{1}-f_{2}\|_{\nu,p}\left(\int_{0}^{\infty}|\bar{g}(x)|^{p}e^{\nu x}dx\right)^{\frac{1}{p}}\\
&+2\left(\frac{p}{\nu q}\right)^{\frac{1}{q}}\|\bar{g}\|_{\nu,p}\|\hat{g}\|_{L^{\infty}}\left(\int_{0}^{\infty}|f_{1}(x)-f_{2}(x)|^{p}e^{\nu x}dx\right)^{\frac{1}{p}}\\
&=4\left(\frac{p}{\nu q}\right)^{\frac{1}{q}}\|\hat{g}\|_{L^{\infty}}\|\bar{g}\|_{\nu,p}\|f_{1}-f_{2}\|_{\nu,p}.
\end{align*}
\noindent Thus,~~$F(t,\cdot)$ is globally Lipschitz on $L^{p}_{\nu}$, which concludes the proof.
\end{proof}

\subsection{Existence and uniqueness of solutions to HJMM equation with locally Lipschitz coefficients}
\label{s:3.3}

\noindent  For each $\nu\in\mathbb{R}$ and $p\geq1$, let $W_{\nu}^{1,p}$ be the weighted Sobolev space defined by
\begin{align*}
W_{\nu}^{1,p}=\{f\in L^{p}_{\nu}~:~Df\in L^{p}_{\nu}\},
\end{align*}
where $Df$ is the first weak derivative of $f$. It is well known that $W_{\nu}^{1,p}$ is a Banach space endowed with the norm
\begin{align*}
\|f\|_{W^{1,p}_{\nu}}=\|f\|_{\nu,p}+\left\|Df\right\|_{\nu,p}.
\end{align*}
Moreover, for $p\geq2$,~$W_{\nu}^{1,p}$ satisfies the $H_{p}$ condition, see \cite{[Kok]} for the proof.

\begin{proposition}\label{P1}
Let $\nu\in\mathbb{R}$ and $p\geq1$. Then there exists a constant $C(\nu,p)>0$ such that
\begin{equation}\label{equa1}
\sup_{x\in[0,\infty)}e^{\nu x}|f(x)|^{p}\leq C^{p}(\nu,p)\|f\|^{p}_{W^{1,p}_{\nu}},\quad f\in W_{\nu}^{1,p}.
\end{equation}
\end{proposition}

\begin{proof}
Take fixed $\nu\in\mathbb{R}$, $p\geq1$ and $f\in W_{\nu}^{1,p}$. Fix $\varepsilon>0$.  Since
\begin{align*}
\int_{0}^{\infty}|f(x)|^{p}e^{\nu x}dx<\infty,
\end{align*}
there exists $x_{0}\in [0,\infty)$ such that $e^{\nu x_{0}}|f(x_{0})|^{p}<\varepsilon$. Consider $x\in[x_{0},\infty)$. Then
\begin{align*}
|f(x)|^{p}&=e^{\nu x_{0}-\nu x}|f(x_{0})|^{p}+e^{-\nu x}\int_{x_{0}}^{x}D\left(|f(x)|^{p}e^{\nu x}\right)dx=e^{\nu x_{0}-\nu x}|f(x_{0})|^{p}\\
&+e^{-\nu x}p\int_{x_{0}}^{x}|f(x)|^{p-1}Df(x)e^{\nu x}dx+e^{-\nu x}\nu\int_{x_{0}}^{x}|f(x)|^{p}e^{\nu x}dx.
\end{align*}

Therefore, we get
\begin{align*}
\sup_{x\in[x_{0},\infty)}e^{\nu x}|f(x)|^{p}\leq\varepsilon+ p\int_{x_{0}}^{\infty}|f(x)|^{p-1}Df(x)e^{\nu x}dx+\nu\|f\|^{p}_{\nu,p}.
\end{align*}

\noindent Using  the H\"older inequality and the Young inequality, we get
\begin{align*}
\int_{x_{0}}^{\infty}|f(x)|^{p-1}&Df(x)e^{\nu x}dx=\int_{x_{0}}^{\infty}|f(x)|^{p-1}Df(x)e^{\frac{(p-1)\nu x}{p}}e^{\frac{\nu x}{p}}dx\\
&\leq\left(\int_{x_{0}}^{\infty}|f(x)|^{p}e^{\nu x}dx\right)^{\frac{p-1}{p}}\left(\int_{x_{0}}^{\infty}|Df(x)|^{p}e^{\nu x}dx\right)^{\frac{1}{p}}\\
&\leq\frac{p-1}{p}\|f\|^{p}_{\nu,p}+\frac{1}{p}\|Df\|^{p}_{\nu,p}.
\end{align*}

\noindent Thus, we infer that

\begin{align*}
\sup_{x\in[x_{0},\infty)}e^{\nu x}|f(x)|^{p}\leq \varepsilon+(p-1)\|f\|^{p}_{\nu,p}+\|Df\|^{p}_{\nu,p}+\nu\|f\|^{p}_{\nu,p}.
\end{align*}
Similarly, we can prove the above inequality for $x\in[0,x_{0})$. Since $\varepsilon>0$ is arbitrary, this concludes the proof.
\end{proof}

\begin{proposition}\label{P3}
Let $\nu>0$ and $p\geq1$. Then for each $f\in W^{1,p}_{\nu}$,
\begin{equation}\label{equa3}
\sup_{x\in[0,\infty)}|f(x)|\leq \|Df\|_{L^{1}}.
\end{equation}
\end{proposition}

\begin{proof}
Fix $\nu>0$, $p\geq1$ and $\varepsilon>0$. Since for any $f\in W^{1,p}_{\nu}$,

\begin{align*}
\int_{0}^{\infty}|f(x)|dx<\infty.
\end{align*}

\noindent Hence, there exists $x_{0}\in [0,\infty)$ such that $|f(x_{0})|<\varepsilon$. Consider  $x\in[x_{0},\infty)$. Then

\begin{align*}
|f(x)-f(x_{0})|=\left|\int_{x_{0}}^{x}Df(x)dx\right|\leq\int_{x_{0}}^{x}\left|Df(x)\right|dx.
\end{align*}

Thus, we get

\begin{align*}
|f(x)|\leq |f(x_{0})|+|f(x)-f(x_{0})|\leq \varepsilon+\int_{x_{0}}^{x}\left|Df(x)\right|dx.
\end{align*}

\noindent Since $\varepsilon$ is arbitrary, we infer that
\begin{align*}
\sup_{x\in[x_{0},\infty)}|f(x)|\leq \left\|Df\right\|_{L^{1}}.
\end{align*}
Similarly, we can prove the above inequality for $x\in[0,x_{0})$. This gives the desired conclusion.
\end{proof}

\begin{lemma}\label{L2}
Let $S=\{S(t)\}_{t\geq0}$ be the family of operators defined in (\ref{shift-semigroup}) for the space $W_{\nu}^{1,p}$. Then $S$ is a contraction type $C_{0}$-semigroup on $W_{\nu}^{1,p}$ and its infinitesimal generator $A$ satisfies
\begin{align*}
D(A)=\left\{f\in W^{1,p}_{\nu} : Df,D^{2}f\in L^{p}_{\nu}\right\},
\end{align*}
\begin{align*}
Af=Df,\quad f\in D(A).
\end{align*}
where $Df$ and $D^{2}f$ denote the first and second weak derivative of $f$ respectively.
\end{lemma}

\noindent For the suitability of our aim, in this section we again assume that $p\geq2$ and $\nu>0$.

\begin{theorem}\label{T1}
Assume that $g~:~[0,\infty)\times[0,\infty)\times\mathbb{R}\rightarrow H$ is a measurable, weakly differentiable mapping with respect to the second and third variables, and $G$ and $F$ are as defined in (\ref{eee1}) and (\ref{ett}). Assume that there exist functions $\bar{g},\hat{g}\in W^{1,p}_{\nu}$ such that
\\

\noindent (i) for all $t\in [0,\infty)$,
\begin{equation}\label{equa35}
\|g(t,x,u)\|_{H}\leq|\bar{g}(x)|,\quad u\in\mathbb{R},\quad x\in [0,\infty),
\end{equation}

\noindent (ii) for all $t\in [0,\infty)$,
\begin{equation}\label{equa36}
\|g(t,x,u)-g(t,x,\rm{v})\|_{H}\leq|\hat{g}(x)|~|u-\rm{v}|,\quad u,\rm{v}\in\mathbb{R},\quad x\in [0,\infty),
\end{equation}

\noindent (iii) for all $t\in [0,\infty)$,
\begin{equation}\label{equa37}
\left\|\frac{\partial}{\partial x}g(t,x,u)\right\|_{H}\leq \left|D\bar{g}(x)\right|,\quad u\in\mathbb{R},\quad x\in [0,\infty),
\end{equation}

\noindent (iv) for all $t\in [0,\infty)$,
\begin{equation}\label{equa38}
\left\|\frac{\partial}{\partial x}g(t,x,u)-\frac{\partial}{\partial x}g(t,x,\rm{v})\right\|_{H}\leq \left|D\hat{g}(x)\right|~|u-\rm{v}|,\quad u,\rm{v}\in\mathbb{R},\quad x\in [0,\infty),
\end{equation}

\noindent (v) there exists $K_{1}>0$ such that for all $t\in [0,\infty)$,
\begin{equation}\label{equa39}
\left\|\frac{\partial}{\partial u}g(t,x,u)\right\|_{H}\leq K_{1},\quad u\in\mathbb{R},\quad x\in [0,\infty),
\end{equation}

\noindent (vi) there exists $K_{2}>0$ such that for all $t\in[0,\infty)$,
\begin{equation}\label{equa40}
\left\|\frac{\partial}{\partial u}g(t,x,u)-\frac{\partial}{\partial \rm{v}}g(t,x,\rm{v})\right\|_{H}\leq K_{2}|u-\rm{v}|,\quad u,\rm{v}\in\mathbb{R},\quad x\in [0,\infty).
\end{equation}

\noindent Then for each $r_{0}\in L^{2}(\Omega,\mathcal{F}_{0},\mathbb{P};W^{1,p}_{\nu})$, there exists a unique $W^{1,p}_{\nu}$-valued mild solution $r$ to equation (\ref{equationn5}) with $r(0)=r_{0}$. Moreover, the solution is a Markov process.
\end{theorem}

\begin{remark}
In the formula of theorem \ref{T1} we denote by $\frac{\partial}{\partial x}g(t,x,u)$ the weak derivative of function $[0,\infty)\ni x\mapsto g(t,x,u)$ when $t$ and $u$ are fixed. Similarly, we denote by $\frac{\partial}{\partial u}g(t,x,u)$ the weak derivative of function $\mathbb{R}\ni u\mapsto g(t,x,u)$ when $t$ and $x$ are fixed.
\end{remark}

In view of Theorem \ref{3theo3},~~the proof of the existence and the uniqueness of solutions follows from Lemma \ref{L2} and the following two lemmata. The proof of the Markov property follows from Theorem \ref{theoremm1}.

\begin{lemma}\label{L3}
Under the all assumptions of Theorem \ref{T1},~~for each $t\geq0$,~~$G(t,\cdot)$ defined in (\ref{eee1}) is a well-defined map from $W^{1,p}_{\nu}$ into $\gamma(H,W^{1,p}_{\nu})$. Moreover,~~for all $t\geq0$,
\\

\noindent (i)
\begin{align*}
\|G(t,f)\|_{\gamma\left(H,W^{1,p}_{\nu}\right)}\leq N \left(\left\|\bar{g}\right\|_{\nu,p}+2\left\|D\bar{g}\right\|_{\nu,p}+2K_{1}\left\|Df\right\|_{\nu,p}\right),\quad f\in W^{1,p}_{\nu}
\end{align*}

\noindent (ii) $G(t,\cdot)$ is locally Lipschitz with Lipschitz constant independent of time $t\geq0$.
\end{lemma}

\begin{proof} \textbf{(i) :}
Fix $t\geq0$ and $f\in W^{1,p}_{\nu}$. Then using (\ref{equa35}) we have
\begin{equation}\label{equa13}
\int_{0}^{\infty}\|g(t,x,f(x)\|^{p}_{H}e^{\nu x}dx\leq\int_{0}^{\infty}\left|\bar{g}(x)\right|^{p}e^{\nu x}dx=\|\bar{g}\|^{p}_{\nu,p}.
\end{equation}

Note that the first weak derivative $Dg(t,\cdot,f(\cdot))$ of function $[0,\infty)\ni x\mapsto g(t,x,f(x))$ is given by, see \cite{[Kok]} for the proof,
\begin{align*}
Dg(t,x,f(x))=\frac{\partial}{\partial x}g(t,x,f(x))+\frac{\partial }{\partial u}g(t,x,f(x))Df(x)\big|_{u=f(x)}.
\end{align*}

Thus, from (\ref{equa37}) and (\ref{equa39})~~we get
\begin{equation}\label{equuuaa}
\left\|Dg(t,x,f(x))\right\|_{H}\leq \left\|\frac{\partial}{\partial x}g(t,x,f(x))\right\|_{H}+\left\|\frac{\partial }{\partial u}g(t,x,f(x))\right\|_{H}\left|Df(x)\right|\leq \left|D\bar{g}(x)\right|+K_{1}\left|Df(x)\right|.
\end{equation}
Therefore, we obtain

\begin{equation}\label{equa14}
\int_{0}^{\infty}\|Dg(t,x,f(x))\|^{p}_{H}e^{\nu x}dx\leq 2^{p}\int_{0}^{\infty}\left|D\bar{g}(x)\right|^{p}e^{\nu x}dx+2^{p}K_{1}^{p}\int_{0}^{\infty}\left|Df(x)\right|^{p}e^{\nu x}dx\leq 2^{p}\left(\left\|D\bar{g}\right\|_{\nu,p}+K_{1}\left\|Df\right\|_{\nu,p}\right)^{p}.
\end{equation}

\noindent It was proven in Theorem 4.1 of \cite{[brzezniak12]} that if $K: H\rightarrow W^{1,p}$,~~$p\in[2,\infty)$,~~is a map defined by
\begin{align*}
K(h)(x)=\langle \kappa(x), h\rangle_{H},\quad x\in[0,\infty),\quad h\in H,
\end{align*}
where $\kappa\in W^{1,p}([0,\infty); H)$,~~then $K\in \gamma\left(H,W^{1,p}\right)$ and for some $N>0$,
\begin{align*}
\|K\|_{\gamma\left(H,W^{1,p}\right)}\leq N\|\kappa\|_{W^{1,p}\left([0,\infty); H\right)}.
\end{align*}

\noindent Therefore, in view of this theorem,~~by (\ref{equa13})~~and~~(\ref{equa14}),~~$G(t,f)$ is a $\gamma$-radonifying operator from $H$ into $W^{1,p}_{\nu}$  and for some $N>0$,

\begin{equation}
\|G(t,f)\|_{\gamma\left(H,W_{\nu}^{1,p}\right)}\leq N \left(\left\|\bar{g}\right\|_{\nu,p}+2\left\|D\bar{g}\right\|_{\nu,p}+2K_{1}\left\|Df\right\|_{\nu,p}\right).
\end{equation}

\noindent \textbf{Proof of (ii)}~~Fix $t\geq0$ and $R>0$ such that $\|f_{1}\|_{W^{1,p}_{\nu}},\|f_{2}\|_{W^{1,p}_{\nu}}\leq R$. Then from Proposition \ref{P1} and (\ref{equa36})~~we get

\begin{align}\label{equa15}
\begin{split}
\int_{0}^{\infty}\|g(t,x,f_{1}(x))-g(t,x,f_{2}(x))\|^{p}_{H}e^{\nu x}dx&\leq \int_{0}^{\infty}\left|\hat{g}(x)\right|^{p}\left|f_{1}(x)-f_{2}(x)\right|^{p}e^{\nu x}dx\\
&\leq \sup_{x\in[0,\infty)}\left|f_{1}(x)-f_{2}(x)\right|^{p}\int_{0}^{\infty}\left|\hat{g}(x)\right|^{p}e^{\nu x}dx\\
&\leq C^{p}(\nu,p)\|\hat{g}\|^{p}_{\nu,p}\|f_{1}-f_{2}\|^{p}_{W^{1,p}_{\nu}}.
\end{split}
\end{align}

Note that the first week derivative of the functions $g(t,\cdot,f_{1}(\cdot))$ and $g(t,\cdot,f_{2}(\cdot))$ are given by respectively
\begin{align}
\begin{split}
&Dg(t,x,f_{1}(x))=\frac{\partial}{\partial x}g(t,x,f_{1}(x))+\frac{\partial }{\partial u}g(t,x,f_{1}(x))Df(x)\big{|}_{u=f_{1}(x)},\\
&Dg(t,x,f_{2}(x))=\frac{\partial}{\partial x}g(t,x,f_{2}(x))+\frac{\partial }{\partial \rm{v}}g(t,x,f_{2}(x))Df(x)\big{|}_{\rm{v}=f_{2}(x)}.
\end{split}
\end{align}
Thus, we obtain
\begin{align*}
&\left\|Dg(t,x,f_{1}(x))-Dg(t,x,f_{2}(x))\right\|_{H}\leq\left\|\frac{\partial}{\partial x}g(t,x,f_{1}(x))-\frac{\partial}{\partial x}g(t,x,f_{2}(x))\right\|_{H}\\
&+\left\|\frac{\partial}{\partial u}g(t,x,f_{1}(x))\right\|_{H}\left|Df_{1}(x)-Df_{2}(x)\right|+\left\|\frac{\partial}{\partial u}g(t,x,f_{1}(x))-\frac{\partial}{\partial \rm{v}}g(t,x,f_{2}(x))\right\|_{H}\left|Df_{2}(x)\right|.
\end{align*}

\noindent It follows from (\ref{equa38}),~(\ref{equa39})~and~(\ref{equa40}) that
\begin{equation}\label{eq:6.3}
\left\|Dg(t,x,f_{1}(x))-Dg(t,x,f_{2}(x))\right\|_{H}\leq \left|D\hat{g}(x)\right|\left|f_{1}(x)-f_{2}(x)\right|+K_{1}\left|Df_{1}(x)-Df_{2}(x)\right|+K_{2}\left|f_{1}(x)-f_{2}(x)\right|\left|Df_{2}(x)\right|.
\end{equation}

\noindent Therefore, by Proposition \ref{P1}, we obtain
\begin{align}\label{equan}
\begin{split}
\int_{0}^{\infty}&\left\|Dg(t,x,f_{1}(x))-Dg(t,x,f_{2}(x))\right|^{p}_{H}e^{\nu x}dx\leq3^{p}\int_{0}^{\infty}\left|D\hat{g}(x)\right|^{p}\left|f_{1}(x)-f_{2}(x)\right|^{p}e^{\nu x}dx\\
&+3^{p}K_{1}^{p}\int_{0}^{\infty}\left|Df_{1}(x)-Df_{2}(x)\right|^{p}e^{\nu x}dx+3^{p}K_{2}^{p}\int_{0}^{\infty}\left|f_{1}(x)-f_{2}(x)\right|^{p}\left|Df_{2}(x)\right|^{p}e^{\nu x}dx\\
&~\leq 3^{p}C^{p}(\nu,p)\left\|f_{1}-f_{2}\right\|^{p}_{W^{1,p}_{\nu}}\left\|D\hat{g}\right\|^{p}_{\nu,p}+3^{p}K_{1}^{p}\left|Df_{1}-Df_{2}\right\|^{p}_{\nu,p}\\
&~~~~~~~~~~~~~~~~~~~~~~~~+3^{p}K_{2}^{p}C^{p} (\nu,p)\left\|f_{1}-f_{2}\right\|^{p}_{W^{1,p}_{\nu}}\left\|Df_{2}\right\|^{p}_{\nu,p}.
\end{split}
\end{align}

Hnece, in view of Theorem 4.1 in \cite{[brzezniak12]}, by estimates (\ref{equa15}) and (\ref{equan}),~~there exist a constant $C(R)>0$ such that
\begin{equation}
\|G(t,f_{1})-G(t,f_{1})\|_{\gamma(H,W^{1,p}_{\nu})}\leq C(R)\left\|f_{1}-f_{2}\right\|^{p}_{W^{1,p}_{\nu}},
\end{equation}
which completes the proof.
\end{proof}

\begin{lemma}\label{L4}
Under the assumption of Theorem \ref{T1},~~for each $t\geq 0$,~~$F(t,\cdot)$ defined in (\ref{ett}) is a well-defined map from $W_{\nu}^{1,p}$ into $W_{\nu}^{1,p}$. Moreover, for all $t\geq0$,
\\

\noindent (i)
\begin{equation}\label{equa21}
\|F(t,f)\|_{W_{\nu}^{1,p}}\leq \left(\frac{p}{\nu q}\right)^{\frac{1}{q}}\|\bar{g}\|^{2}_{\nu,p}+3\left(\frac{p}{\nu q}\right)^{\frac{1}{q}}\|\bar{g}\|_{\nu,p}\left(\|D\bar{g}\|_{\nu,p}+K_{1}\|Df\|_{\nu,p}\right)+3C(\nu,p)\|\bar{g}\|_{W^{1,p}_{\nu}}\|\bar{g}\|_{\nu,p},\quad f\in W_{\nu}^{1,p}
\end{equation}

\noindent (ii) $F(t,\cdot)$ is locally Lipschitz with Lipschitz constant independent of time $t\geq0$.
\end{lemma}

\begin{proof}\textbf{(i) :}
Fix $t\geq0$ and $f\in W_{\nu}^{1,p}$. Then by the chain rule and the Cauchy-Schwarz inequality,~~we obtain (with $DF(t,f)$ being the first weak derivative of $F(t,f)$ w.r.t $x$)
\begin{align*}
\left|DF(t,f)(x)\right|&=\left|\left\langle Dg(t,x,f(x)), \int_{0}^{x}g(t,y,f(y))dy\right\rangle_{H}+\left\langle g(t,x,f(x)), g(t,x,f(x))\right\rangle_{H}\right|\\
&\leq \left\|D(g(t,x,f(x)))\right\|_{H}\int_{0}^{\infty}\|g(t,x,f(x))\|_{H}dx+\|g(t,x,f(x))\|_{H}\|g(t,x,f(x))\|_{H}.
\end{align*}

\noindent It follows from Proposition \ref{prr1} and (\ref{equa35}) that
\begin{align*}
\left|DF(t,f)(x)\right|&\leq \left\|Dg(t,x,f(x)\right\|_{H}\int_{0}^{\infty}\left|\bar{g}(x)\right|dx+\left|\bar{g}(x)\right|^{2}\\
&\leq \left(\frac{p}{\nu q}\right)^{\frac{1}{q}}\left\|\bar{g}\right\|_{\nu,p}\left\|Dg(t,x,f(x))\right\|_{H}+\left|\bar{g}(x)\right|^{2}.
\end{align*}

\noindent Thus, by inequality (\ref{equuuaa}),~~we get
\begin{align*}
\left|DF(t,f)(x)\right|\leq \left(\frac{p}{\nu q}\right)^{\frac{1}{q}}\|\bar{g}\|_{\nu,p}\left|D\bar{g}(x)\right|+K_{1}\left(\frac{p}{\nu q}\right)^{\frac{1}{q}}\left\|\bar{g}\right\|_{\nu,p}\left|Df(x)\right|+\left|\bar{g}(x)\right|^{2}.
\end{align*}

Hence, by Proposition \ref{prr1} and Proposition \ref{P1},~~we obtain
\begin{equation}\label{eeeeee1}
\|DF(t,f)\|_{\nu,p}\leq 3\left(\frac{p}{\nu q}\right)^{\frac{1}{q}}\|\bar{g}\|_{\nu,p}\left\|D\bar{g}\right\|_{\nu,p}+3\left(\frac{p}{\nu q}\right)^{\frac{1}{q}}\left\|\bar{g}\right\|_{\nu,p}K_{1}\left\|Df\right\|_{\nu,p}+3C(\nu,p)\|\bar{g}\|_{\nu.p}\|\bar{g}\|_{W^{1.p}_{\nu}}.
\end{equation}
Therefore,~~by estimates (\ref{eeqq9}) and (\ref{eeeeee1}),~~$F(t,\cdot)$ is well-defined and
\begin{align*}
\|F(t,f)\|_{W^{1,p}_{\nu}}\leq \left(\frac{p}{\nu q}\right)^{\frac{1}{q}}\|\bar{g}\|^{2}_{\nu,p}+3\left(\frac{p}{\nu q}\right)^{\frac{1}{q}}\|\bar{g}\|_{\nu,p}\left(\|D\bar{g}\|_{\nu,p}+K_{1}\|Df\|_{\nu,p}\right)+3C(\nu,p)\|\bar{g}\|_{W^{1,p}_{\nu}}\|\bar{g}\|_{\nu,p}.
\end{align*}

\noindent \textbf{Proof of (ii) :} Fix $t\geq0$  and $R>0$ such that $\|f_{1}\|_{W^{1,p}_{\nu}},\|f_{2}\|_{W^{1,p}_{\nu}}\leq R$. Then by the Cauchy-Schwarz inequality,~~we have

\begin{align*}
\left|F(t,f_{1})(x)-F(t,f_{2})(x)\right|&=\left|\left\langle g(t,x,f_{1}(x)),\int_{0}^{x}g(t,y,f_{1}(y))dy\right\rangle_{H}-\left\langle g(t,x,f_{2}(x)),\int_{0}^{x}g(t,y,f_{2}(y))dy\right\rangle_{H}\right|\\
&\leq\left|\left\langle g(t,x,f_{1}(x))-g(t,x,f_{2}(x)),\int_{0}^{x}g(t,y,f_{2}(y))dy\right\rangle_{H}\right|\\
&+\left|\left\langle g(t,x,f_{1}(x)),\int_{0}^{x}\left[g(t,x,f_{1}(x))-g(t,y,f_{2}(y))\right]dy\right\rangle_{H}\right|\\
&\leq \left\|g(t,x,f_{1}(x))-g(t,x,f_{2}(x))\right\|_{H}\int_{0}^{\infty}\left\|g(t,x,f_{2}(x))\right\|_{H}dx\\
&+\left\|g(t,x,f_{1}(x))\right\|_{H}\int_{0}^{\infty}\left\|g(t,x,f_{1}(x))-g(t,x,f_{2}(x))\right\|dx.
\end{align*}

\noindent By (\ref{equa35}) and (\ref{equa36}),~~we obtain

\begin{align*}
\left|F(t,f_{1})(x)-F(t,f_{2})(x)\right|\leq \left|\hat{g}(x)\right|\left|f_{1}(x)-f_{2}(x)\right|\int_{0}^{\infty}\left|\bar{g}(x)\right|dx+\left|\bar{g}(x)\right|\int_{0}^{\infty}\left|\hat{g}(x)\right|\left|f_{1}(x)-f_{2}(x)\right|dx.
\end{align*}

\noindent It follows from Proposition \ref{prr1} and Proposition \ref{P3} that
\begin{align*}
\left|F(t,f_{1})(x)-F(t,f_{2})(x)\right|\leq \left(\frac{p}{\nu q}\right)^{\frac{1}{q}}\left\|\bar{g}\right\|_{\nu,p}\left|\hat{g}(x)\right|\left|f_{1}(x)-f_{2}(x)\right|+\left(\frac{p}{\nu q}\right)^{\frac{1}{q}}\left\|f_{1}-f_{1}\right\|_{\nu,p}\left\|D\hat{g}\right\|_{L^{1}}\left|\bar{g}(x)\right|.
\end{align*}

\noindent Therefore, by Proposition \ref{P1},~~we infer that
\begin{equation}\label{ett1}
\|F(t,f_{1})-F(t,f_{2})\|_{\nu,p}\leq 2C(\nu,p)\left(\frac{p}{\nu q}\right)^{\frac{1}{q}}\left\|\bar{g}\right\|_{\nu,p} \left\|\hat{g}\right\|_{W^{1,p}_{\nu}}\left\|f_{1}-f_{2}\right\|_{\nu,p}+2\left(\frac{p}{\nu q}\right)^{\frac{1}{q}}\left\|f_{1}-f_{1}\right\|_{\nu,p}\left\|D\hat{g}\right\|_{L^{1}}\left\|\bar{g}\right\|_{\nu,p}.
\end{equation}

\noindent By the chain rule,~~we get
\begin{align*}
DF(t,f_{1})(x)-DF(t,f_{2})(x)&=\left\langle Dg(t,x,f_{1}(x)),\int_{0}^{x}g(t,y,f_{1}(y))dy\right\rangle_{H}+\left\langle g(t,x,f_{1}(x)),g(t,x,f_{1}(x))\right\rangle_{H}\\
&\quad -\left\langle Dg(t,x,f_{2}(x)),\int_{0}^{x}g(t,y,f_{2}(y))dy\right\rangle_{H}-\left\langle g(t,x,f_{2}(x)),g(t,x,f_{2}(x))\right\rangle_{H}\\
&=\left\langle Dg(t,x,f_{1}(x)),\int_{0}^{x}(g(t,y,f_{1}(y))-g(t,y,f_{2}(y)))dy\right\rangle_{H}\\
&\quad +\left\langle Dg(t,x,f_{1}(x))-Dg(t,x,f_{2}(x)),\int_{0}^{\infty}g(t,x,f_{2}(x))dx\right\rangle_{H}\\
&\quad +\left\langle g(t,x,f_{1}(x)),[g(t,x,f_{1}(x))-g(t,x,f_{2}(x))]\right\rangle_{H}\\
&\quad+\left\langle g(t,x,f_{1}(x))-g(t,x,f_{2}(x)),g(t,x,f_{2}(x))\right\rangle_{H}.
\end{align*}

\noindent Thus, using the Cauchy-Schwarz inequality~~we get
\begin{align*}
\left|DF(t,f_{1})(x)-DF(t,f_{2})(x)\right|&\leq \left\|Dg(t,x,f_{1}(x))\right\|_{H}\int_{0}^{\infty}\left\|g(t,x,f_{1}(x))-g(t,x,f_{2}(x))\right\|_{H}dx\\
&+\left\|Dg(t,x,f_{1}(x))-Dg(t,x,f_{2}(x)\right\|_{H}\int_{0}^{\infty}\left\|g(t,x,f_{2}(x))\right\|_{H}dx\\
&+2\left\|g(t,x,f_{1}(x))\right\|_{H}\left\|g(t,x,f_{1}(x))-g(t,x,f_{2}(x))\right\|_{H}.
\end{align*}

\noindent It follows from (\ref{equa35}) and (\ref{equa36}) that

\begin{align*}
&\left|DF(t,f_{1})(x)-DF(t,f_{2})(x)\right|\leq \left\|Dg(t,x,f_{1}(x))\right\|_{H}\int_{0}^{\infty}\left|\hat{g}(x)\right|\left|f_{1}(x)-f_{2}(x)\right|dx\\
&+\left\|Dg(t,x,f_{1}(x))-Dg(t,x,f_{2}(x)\right\|_{H}\int_{0}^{\infty}\left|\bar{g}(x)\right|dx+2\left|\bar{g}(x)\right|\left|\hat{g}(x)\right|\left|f_{1}(x)-f_{2}(x)\right|\\
&\leq\left\| Dg(t,x,f_{1}(x))\right\|_{H}\sup_{x\in[0,\infty)}\left|\hat{g}(x)\right|\int_{0}^{\infty}\left|f_{1}(x)-f_{2}(x)\right|dx+\left\|Dg(t,x,f_{1}(x))-Dg(t,x,f_{2}(x)\right\|_{H}\int_{0}^{\infty}\left|\bar{g}(x)\right|dx\\
&~~~~~~~~~~~~~~~~~~~~~~~~~~~~~~~~~~~~~~~~~~~~~~~~~~~~~~~~~~~~~~~~~~~~~~~~~~+2\left|\bar{g}(x)\right|\left|\hat{g}(x)\right|\left|f_{1}(x)-f_{2}(x)\right|.
\end{align*}

\noindent By Proposition \ref{prr1}~~and~~Proposition \ref{P3},~~we obtain

\begin{align*}
\left|DF(t,f_{1})(x)-DF(t,f_{2})(x)\right|&\leq \left(\frac{p}{\nu q}\right)^{\frac{1}{q}}\left\|D\hat{g}\right\|_{L^{1}}\left\|f_{1}-f_{2}\right\|_{\nu,p}\left\| Dg(t,x,f_{1}(x))\right\|_{H}\\
&+\left(\frac{p}{\nu q}\right)^{\frac{1}{q}}\left\|\bar{g}\right\|_{\nu,p}\left\|Dg(t,x,f_{1}(x))-Dg(t,x,f_{2}(x)\right\|_{H}\\
&+2\left|\bar{g}(x)\right|\left|\hat{g}(x)\right|\left|f_{1}(x)-f_{2}(x)\right|.
\end{align*}

\noindent It follows from inequalities (\ref{equuuaa}) and (\ref{eq:6.3}) that
\begin{align*}
\left|DF(t,f_{1})(x)-DF(t,f_{2})(x)\right|&\leq \left(\frac{p}{\nu q}\right)^{\frac{1}{q}}\left\|D\hat{g}\right\|_{L^{1}}\left\|f_{1}-f_{2}\right\|_{\nu,p}\left|D\bar{g}(x)\right|+K_{1}\left(\frac{p}{\nu q}\right)^{\frac{1}{q}}\left\|D\hat{g}\right\|_{L^{1}}\left\|f_{1}-f_{2}\right\|_{\nu,p}\left|Df_{1}(x)\right|\\
&+\left(\frac{p}{\nu q}\right)^{\frac{1}{q}}\left\|\bar{g}\right\|_{\nu,p}\left|D\hat{g}(x)\right|\left|f_{1}(x)-f_{2}(x)\right|+K_{1}\left(\frac{p}{\nu q}\right)^{\frac{1}{q}}\left\|\bar{g}\right\|_{\nu,p}\left|Df_{1}(x)-Df_{2}(x)\right|\\
&+ K_{2}\left(\frac{p}{\nu q}\right)^{\frac{1}{q}}\left\|\bar{g}\right\|_{\nu,p}\left|f_{1}(x)-f_{2}(x)\right|\left|Df_{2}(x)\right|+2\left|\bar{g}(x)\right|\left|\hat{g}(x)\right|\left|f_{1}(x)-f_{2}(x)\right|.
\end{align*}

\noindent Therefore, we obtain
\begin{align*}
\left\|DF(t,f_{1})-DF(t,f_{2})(x)\right\|_{\nu,p}&\leq \left(\frac{p}{\nu q}\right)^{\frac{1}{q}}\left\|D\hat{g}\right\|_{L^{1}}\left\|f_{1}-f_{2}\right\|_{\nu,p}\left(\int_{0}^{\infty}\left|D\bar{g}(x)\right|^{p}e^{\nu x}dx\right)^{\frac{1}{p}}\\
&+K_{1}\left(\frac{p}{\nu q}\right)^{\frac{1}{q}}\left\|D\hat{g}\right\|_{L^{1}}\left\|f_{1}-f_{2}\right\|_{\nu,p}\left(\int_{0}^{\infty}\left|Df_{1}(x)\right|^{p}e^{\nu x}dx\right)^{\frac{1}{p}}\\
&+\left(\frac{p}{\nu q}\right)^{\frac{1}{q}}\left\|\bar{g}\right\|_{\nu,p}\left(\int_{0}^{\infty}\left|D\hat{g}(x)\right|^{p}\left|f_{1}(x)-f_{2}(x)\right|^{p}e^{\nu x}dx\right)^{\frac{1}{p}}\\
&+K_{1}\left(\frac{p}{\nu q}\right)^{\frac{1}{q}}\left\|\bar{g}\right\|_{\nu,p}\left(\int_{0}^{\infty}\left|Df_{1}(x)-Df_{2}(x)\right|^{p}e^{\nu x}dx\right)^{\frac{1}{p}}\\
&+K_{2}\left(\frac{p}{\nu q}\right)^{\frac{1}{q}}\left\|\bar{g}\right\|_{\nu,p}\left(\int_{0}^{\infty}\left|f_{1}(x)-f_{2}(x)\right|^{p}\left|Df_{2}(x)\right|^{p}e^{\nu x}dx\right)^{\frac{1}{p}}\\
&+2\left(\int_{0}^{\infty}\left|\bar{g}(x)\right|^{p}\left|\hat{g}(x)\right|^{p}\left|f_{1}(x)-f_{2}(x)\right|^{p}e^{\nu x}dx\right)^{\frac{1}{p}}.
\end{align*}

\noindent It follows from Proposition \ref{P1} that
\begin{equation}\label{equa34}
\begin{split}
\big\|DF(t,f_{1})-DF(t,&f_{2})(x)\big\|_{\nu,p}\leq \left(\frac{p}{\nu q}\right)^{\frac{1}{q}}\left\|D\hat{g}\right\|_{L^{1}}\left\|f_{1}-f_{2}\right\|_{\nu,p}\left\|D\bar{g}\right\|_{\nu,p}+K_{1}\left(\frac{p}{\nu q}\right)^{\frac{1}{q}}\left\|D\hat{g}\right\|_{L^{1}}\left\|f_{1}-f_{2}\right\|_{\nu,p}\left\|Df_{1}\right\|_{\nu,p}\\
&+C(v,P)\left(\frac{p}{\nu q}\right)^{\frac{1}{q}}\left\|\bar{g}\right\|_{\nu,p}\left\|D\hat{g}\right\|_{\nu,p}\left\|f_{1}-f_{2}\right\|_{W^{1,p}_{\nu}}+ K_{1}\left(\frac{p}{\nu q}\right)^{\frac{1}{q}}\left\|\bar{g}\right\|_{\nu,p}\left\|Df_{1}-Df_{2}\right\|_{\nu,p}\\
&+C(v,P)K_{2}\left(\frac{p}{\nu q}\right)^{\frac{1}{q}}\left\|\bar{g}\right\|_{\nu,p}\left\|f_{1}-f_{2}\right\|_{W^{1,p}_{\nu}}\left\|f^{'}_{2}\right\|_{\nu,p}+2C^{2}(\nu,p)\left\|f_{1}-f_{2}\right\|_{W^{1,p}_{\nu}}\left\|\hat{g}\right\|_{W^{1,p}_{\nu}}\left\|\hat{g}\right\|_{\nu,p}.
\end{split}
\end{equation}
Therefore,~~by inequalities (\ref{ett1}) and (\ref{equa34}),~~there exists $C(R)>0$ such that

\begin{align*}
\|F(t,f_{1})-F(t,f_{2})\|_{W^{1,p}_{\nu}}\leq C(R) \|f_{1}-f_{2}\|_{W^{1,p}_{\nu}},
\end{align*}

which concludes the proof.
\end{proof}

\subsection{Existence and Uniqueness of Invariant Measures for  Equation (\ref{equationn5})}
\label{s:3.4}

\noindent In the section we consider that the coefficients $G$ and $F$ of equation (\ref{equationn5}) are time independent. The following corollary is a natural consequence of Theorem \ref{theo1}.

\begin{corollary}\label{corr1}
Let $g:[0,\infty)\times\mathbb{R}\rightarrow H$ be a measurable function such that there are functions $\bar{g}\in L^{p}_{\nu}$ and $\hat{g}\in L^{p}_{\nu}\cap L^{\infty}$ such that
\begin{equation}\label{eeeqqq1}
\|g(x,u)\|_{H}\leq|\bar{g}(x)|,\quad u\in\mathbb{R},\quad x\in [0,\infty)
\end{equation}
and
\begin{equation}\label{eeeqqq2}
\|g(x,u)-g(t,x,\rm{v})\|_{H}\leq|\hat{g}(x)|~|u-\rm{v}|,\quad u,\rm{v}\in\mathbb{R},\quad x\in [0,\infty).
\end{equation}
Assume that  maps $G$  and $F$  are defined respectively by
\begin{equation}
G(f)[h](x)=\langle g(x,f(x)),h\rangle_{H},\quad f\in L^{p}_{\nu},\quad x\in[0,\infty),\quad h\in H.
\end{equation}
\begin{equation}
F(f)(x)=\langle g(x,f(x)),\int_{0}^{x}g(t,f(y))dy\rangle_{H},\quad f\in L^{p}_{\nu},\quad x\in[0,\infty).
\end{equation}
Then for each $r_{0}\in L^{2}(\Omega,\mathcal{F}_{0},\mathbb{P};L^{p}_{\nu})$, there exists a unique $L^{p}_{\nu}$-valued mild solution $r$ to equation (\ref{equationn5}) with $r(0)=r_{0}$ and the solution is a Markov process.
\end{corollary}

\begin{theorem}
Let us assume that all assumptions of Corollary \ref{corr1} are satisfied. If $\frac{1}{p}+\frac{1}{q}=1$ and
\begin{equation}\label{equu4}
2\left(\frac{p}{\nu q}\right)^{\frac{1}{q}}||\hat{g}||_{L^{\infty}}||\bar{g}||_{\nu,p}+(p-1)N^{2}||\hat{g}||^{2}_{L^{\infty}}<\frac{\nu}{2},
\end{equation}
then equation (\ref{equationn5}) has a unique invariant measure in the spaces $L^{p}_{\nu}$.
\end{theorem}

 \begin{proof}
 In the view of Theorem \ref{theoremm2} it is enough to check that there exist a constant $\omega>0$ and a natural number $n_{0}\in\mathbb{N}$ such that for all $f_{1},f_{2}\in L^{p}_{\nu}$ and $n\geq n_{0}$,

 \begin{equation}\label{equu5}
 [A_{n}(f_{1}-f_{2})+F(f_{1})-F(f_{2}),f_{1}-f_{2}]+\frac{K_{2}(p)}{p}||G(f_{1})-G(f_{2})||^{2}_{\gamma(H,L^{p}_{\nu})}\leq-\omega||f_{1}-f_{2}||_{\nu,p}^{2},
 \end{equation}
 where $[\cdot,\cdot]$ is the semi-inner product on $L^{p}_{\nu}\times L^{p}_{\nu}$ which is given, see \cite{[brzezniak8]}, by
 \begin{align*}
 [f,g]=||g||^{2-p}_{\nu,p}\int_{0}^{\infty}f(x)g(x)|g(x)|^{p-2}e^{\nu x}dx,\quad f,g\in L^{p}_{\nu}.
 \end{align*}

\noindent We will prove the theorem in the following few steps.

\noindent \textbf{Step 1 :} Fix $f_{1},f_{2}\in L^{p}_{\nu}$. Then by the Cauchy-Schwarz inequality~~we obtain
\begin{align*}
\big|F(f_{1})(x)-F(f_{2})(x)\big|&=\left|\left\langle g(x,f_{1}(x)), \int_{0}^{x}g(y,f_{1}(y))dy\right\rangle_{H}-\left\langle g(x,f_{2}(x)),\int_{0}^{x}g(y,f_{2}(y))dy\right\rangle_{H}\right|\\
&=\left|\left\langle g(x,f_{1}(x)),\int_{0}^{x}[g(y,f_{1}(y))-g(y,f_{2}(y))]dy\right\rangle_{H}\right|\\
&\quad\quad +\left|\left\langle g(x,f_{1}(x))-g(x,f_{2}(x)), \int_{0}^{x}g(y,f_{2}(y))dy\right\rangle_{H}\right|\\
&\leq \left\|g(x,f_{1}(x))\right\|_{H} \int_{0}^{\infty}\left\|g(y,f_{1}(y))-g(y,f_{2}(y))\right\|_{H}dy\\
&\quad\quad +\left\|g(x,f_{1}(x))-g(x,f_{2}(x))\right\|_{H}\int_{0}^{\infty}\left\|g(y,f_{2}(y))\right\|_{H}dy.
\end{align*}

 \noindent Hence it follows  from Proposition \ref{prr1}, (\ref{eeeqqq1}) and  (\ref{eeeqqq2}) that

\begin{align*}
\left|F(f_{1})(x)-F(f_{2})(x)\right|\leq\left(\frac{p}{\nu q}\right)^{\frac{1}{q}}\left\|\hat{g}\right|_{L^{\infty}}||f_{1}-f_{2}||_{\nu,p}|\bar{g}(x)|+ \left(\frac{p}{\nu q}\right)^{\frac{1}{q}}||\bar{g}||_{\nu,p}|\hat{g}(x)|~|f_{1}(x)-f_{2}(x)|.
\end{align*}

\noindent Thus, we obtain

\begin{align*}
[F(f_{1})-F(f_{2}),f_{1}-f_{2}]&\leq \left(\frac{p}{\nu q}\right)^{\frac{1}{q}}||\hat{g}||_{L^{\infty}}||f_{1}-f_{2}||^{3-p}_{\nu,p} \int_{0}^{\infty}|\bar{g}(x)|~|f_{1}(x)-f_{2}(x)|^{p-1}e^{\nu x}dx\\
&+\left(\frac{p}{\nu q}\right)^{\frac{1}{q}}||\bar{g}||_{\nu,p}\|f_{1}-f_{2}\|^{2-p}_{\nu,p} \int_{0}^{\infty}|\hat{g}(x)|~|f_{1}(x)-f_{2}(x)|^{p}e^{\nu x}dx.
\end{align*}

\noindent By the H\"older inequality,~~we get
\begin{align*}
\int_{0}^{\infty}|\bar{g}(x)|~|f_{1}(x)-f_{2}(x)|^{p-1}&e^{\nu x}dx=\int_{0}^{\infty}|\bar{g}(x)|~|f_{1}(x)-f_{2}(x)|^{p-1}e^{\nu x\left(\frac{p-1}{p}\right)} e^{\frac{\nu x}{p}}dx\\
&\leq\left(\int_{0}^{\infty}|\bar{g}(x)|^{p}e^{\nu x}dx\right)^{\frac{1}{p}}\left(\int_{0}^{\infty}|f_{1}(x)-f_{2}(x)|^{p}e^{\nu x}dx\right)^{\frac{p-1}{p}}\\
&=\|\bar{g}\|_{\nu,p}\|f_{1}-f_{2}\|^{p-1}_{\nu,p}.
\end{align*}
Since
\begin{align*}
\int_{0}^{\infty}|\hat{g}(x)|~|f_{1}(x)-f_{2}(x)|^{p}e^{\nu x}dx\leq \|\hat{g}\|_{L^{\infty}}\|f_{1}-f_{2}\|^{p}_{\nu,p}.
\end{align*}

\noindent we deduce that

\begin{equation}\label{ee4}
[F(f_{1})-F(f_{2}),f_{1}-f_{2}]\leq \omega_{1}\|f_{1}-f_{2}\|^{2}_{\nu,p},
\end{equation}
where $\omega_{1}=2\left(\frac{p}{\nu q}\right)^{\frac{1}{q}}\left\|\hat{g}\right\|_{L^{\infty}}\left\|\bar{g}\right\|_{\nu,p}$.
\\

\noindent \textbf{Step 2 :} Define
\begin{align*}
P(t)=e^{\frac{\nu t}{p}}S(t),\quad t\geq0.
\end{align*}
Since $||S(t)||_{L(L^{p}_{\nu})}\leq e^{\frac{-\nu t}{p}}$,~~$\|P(t)\|_{L(L^{p}_{\nu})}\leq1$. Therefore $P$ is a contraction $C_{0}$-semigroup on $L^{p}_{\nu}$ with the infinitesimal generator given by
\begin{align*}
B=\frac{\nu I}{p}+A.
\end{align*}

\noindent From Theorem $4.3$ (b) in \cite{[Pazy]}~ $B$ is dissipative and hence for all $f\in D(B)$ and $f^{*}\in F(f)$, $\langle Bf,f^{*}\rangle\leq 0$, where $F(f)=\left\{f^{*}\in (L^{p}_{\nu})^{*}~:~\langle f,f^{*}\rangle=||f||^{2}=||f^{*}||^{2}\right\}$. By the definition of the semi-inner product $[\cdot,\cdot]$,~~see Definition \ref{def-semi inner product},~~$[f,g]=\langle f,g^{*}\rangle$,~~$g^{*}\in (L^{p}_{\nu})^{*}$ and $\langle g, g^{*}\rangle=||g||^{2}$, i.e. $g^{*}\in (L^{p}_{\nu})^{*}$. Therefore, $\langle Bf, f^{*}\rangle=[Bf,f]$. Hence $[Bf,f]\leq0$.
\begin{align*}
A_{n}=nA(nI-A)^{-1}=n\left(B-\frac{\nu}{p}I\right)\left(nI+\frac{\nu}{p}I-B\right)^{-1}.
\end{align*}
Let $\omega_{2}=\frac{-\nu}{p}$ and $k=n-\omega_{2}$. Then
\begin{align*}
A_{n}=(\omega^{2}+k\omega_{2})(kI-B)^{-1}+\left(1+\frac{\omega_{2}}{k}\right) B_{k}.
\end{align*}

\noindent Therefore, we get
\begin{align*}
[A_{n}f,f]\leq (\omega_{2}^{2}+k\omega_{2})\|(kI-B)^{-1}\|~\|f\|^{2}_{\nu,p}.
\end{align*}
By Theorem 3.1 in \cite{[Pazy]},~~$\|(kI-B)^{-1}\|\leq \frac{1}{k}$. Thus,
\begin{align*}
[A_{n}f,f]\leq \frac{\omega_{2}^{2}+k\omega_{2}}{k}\|f\|^{2}_{\nu,p}
\end{align*}
and since $k=n-\omega_{2}$, we infer that

\begin{align*}\label{eqqq7}
[A_{n}f,f]\leq \frac{n\omega_{2}}{n-\omega_{2}}||f||^{2}_{\nu,p}.
\end{align*}

\noindent Therefore, we obtain

\begin{equation}\label{eqqq5}
[A_{n}(f_{1}-f_{2}),f_{1}-f_{2}]\leq \frac{n\omega_{2}}{n-\omega_{2}}\|f_{1}-f_{2}\|^{2}_{\nu,p},\quad f_{1},f_{2}\in L^{p}_{\nu}.
\end{equation}

\noindent \textbf{Step 3 :} By inequalities (\ref{eee10}) and (\ref{ee5}),~~we get
\begin{equation}\label{eeeee}
\frac{K_{2}(p)}{p}||G(f_{1})-G(f_{2})||^{2}_{\gamma(H,L^{p}_{\nu})}\leq \omega_{3}\|f_{1}-f_{2}\|^{2}_{\nu,p},
\end{equation}
where $\omega_{3}=(p-1)N^{2}\left\|\hat{g}\right\|^{2}_{L^{\infty}}$. Taking into account inequalities (\ref{ee4}), (\ref{eqqq5}) and (\ref{eeeee})~~we obtain
\begin{align*}
[A_{n}(f_{1}-f_{2})+F(f_{1})-F(f_{2}),f_{1}-f_{2}]+\frac{K_{p}}{p}\|G(f_{1})-G(f_{2})\|^{2}_{\gamma(H,L^{p}_{\nu}}\leq \left(\omega_{1}+\omega_{3}+\frac{n\omega_{2}}{n-\omega_{2}}\right)~\|f_{1}-f_{2}\|^{2}_{\nu,p}.
\end{align*}

\noindent Since $\omega_{1}+\omega_{3}+\frac{n\omega_{2}}{n-\omega_{2}}\rightarrow \omega_{1}+\omega_{2}+\omega_{3}<0$ by (\ref{equu4}),~~we infer that $\exists~n_{0}\in\mathbb{N}$ such that $\forall~\omega\in (0,-(\omega_{1}+\omega_{2}+\omega_{3}))$,
\begin{align*}
\omega_{1}+\omega_{3}+\frac{n\omega_{2}}{n-\omega_{2}}\leq -\omega,\quad n\geq n_{0}.
\end{align*}

\noindent Therefore (\ref{equu5}) holds for any $\omega\in(0,-(\omega_{1}+\omega_{2}+\omega_{3}))$ and $n\geq\frac{\omega_{2}(\omega+\omega_{1}+\omega_{3})}{\omega_{1}+\omega_{2}+\omega_{3}+\omega}$. This finishes the proof.
\end{proof}

\begin{remark}
It would be interesting to investigate the existence and the uniqueness of invariant measures for equation (\ref{equationn5}) in the spaces $W^{1,p}_{\nu}$.
\end{remark}

\begin{remark}
In the subsequent paper \cite{[brzezniak13]}  we will study the existence, uniqueness and ergodic properties of solutions to a nonlinear modification of HJMM equation, proposed by \cite{[Goldys]}, in the weighted $L^{p}_{\nu}$ spaces.
\end{remark}

\section*{Acknowledgements}
We would like to thank Prof Lutz Weis for his hospitality during our visit in Karlsruhe Institute of Technology (KIT). Also second name author would like to thank Ministry of National Education of Republic of Turkey for funding him during his PhD. We would also like to thank to Prof Ben Goldys for stimulating discussion on various parts of this work.

\end{document}